\documentclass[12pt, preprint]{aastex}
\usepackage{lscape}
\usepackage{subfigure}
\usepackage{graphicx}
\usepackage{multirow}
\usepackage{latexsym}
\usepackage{amssymb}
\usepackage{epsf}

\newcommand{\kms}{km s$^{-1}$ }
\newcommand{\kmsn}{km s$^{-1}$}
\newcommand{\Msu}{$M_{\odot}$ }
\newcommand{\Msun}{$M_{\odot}$}
\newcommand{\degs}{$^{\circ}$ }
\newcommand{\degn}{$^{\circ}$}

\newcommand{\his}{{\rm H\,}{{\sc i }}}
\newcommand{\hiis}{{\rm H\,}{{\sc ii }}}

\newcommand{\cii}{{\rm C\,}{{\sc ii}}}

\newcommand{\CO}{$^{13}$CO }
\newcommand{\COJ}{$^{13}$CO J =  $ 1 \rightarrow 0 $ }
\newcommand{\COT}{$^{12}$CO }
\newcommand{\COTJ}{$^{12}$CO J =  $ 1 \rightarrow 0 $ }

\begin{document}	\title{Physical properties and Galactic Distribution of molecular clouds identified in the Galactic Ring Survey}

\author{Julia Roman-Duval\altaffilmark{1, 2},  James M. Jackson\altaffilmark{1}, Mark Heyer\altaffilmark{3}, Jill Rathborne\altaffilmark{4, 5}, Robert Simon\altaffilmark{6}}
\altaffiltext{1}{Institute for Astrophysical Research at Boston University, 725 Commonwealth Avenue, Boston MA 02215; jduval@bu.edu, jackson@bu.edu}
\altaffiltext{2}{Space Telescope Science Institute, 3700 San Martin Drive, Baltimore, MD 21218; duval@stsci.edu}
\altaffiltext{3}{Department of Astronomy, University of Massachusetts, Amherst, MA 01003-9305, heyer@astro.umass.edu}
\altaffiltext{4}{Departamento de Astronomia, Universidad de Chile, Santiago, Chile}
\altaffiltext{5}{CSIRO Astronomy and Space Science, PO Box 76, Epping NSW 1710, Australia}
\altaffiltext{6}{Physikalisches Instit\"{u}t, Universit\"{a}t zu K\"{o}ln, 50937 K\"{o}ln, Germany, simonr@ph1.uni-koeln.de}

\begin{abstract}
\indent We derive the physical properties of 580 molecular clouds based on their \COT and \CO line emission detected in the University of Massachusetts-Stony Brook (UMSB) and Galactic Ring surveys. We provide a range of values of the physical properties of molecular clouds, and find a power-law correlation between their radii and masses, suggesting that the fractal dimension of the ISM is around 2.36. This relation, M $=$ (228$\pm$18) R$^{2.36\pm0.04}$, allows us to derive masses for an additional 170 GRS molecular clouds not covered by the UMSB survey. We derive the Galactic surface mass density of molecular gas and examine its spatial variations throughout the Galaxy. We find that the azimuthally averaged Galactic surface density of molecular gas peaks between Galactocentric radii of 4 and 5 kpc. Although the Perseus arm is not detected in molecular gas, the Galactic surface density of molecular gas is enhanced along the positions of the Scutum-Crux and Sagittarius arms. This may indicate that molecular clouds form in spiral arms and are disrupted in the inter-arm space. Last, we find that the CO excitation temperature of molecular clouds decreases away from the Galactic center, suggesting a possible decline in the star formation rate with Galactocentric radius. There is a marginally significant enhancement in the CO excitation temperature of molecular clouds at a Galactocentric radius of about 6 kpc, which in the longitude range of the GRS corresponds to the Sagittarius arm. This temperature increase could be associated with massive star formation in the Sagittarius spiral arm.

\end{abstract}

\maketitle

\section{Introduction} \label{introduction}
\indent In the past forty years, there has been a considerable effort to establish the physical properties of molecular clouds, to understand how turbulence, gravity and magnetic fields shape their complex structure, how stars are born from their collapse, and if/how they relate to the spiral structure of the Milky Way.  Several questions however remain unanswered. In particular, it is not clear how the spiral structure of our Galaxy relates to the formation and distribution of molecular clouds. This lack of evidence for a spiral structure traced by molecular clouds comes from a confused view of our own Galaxy, and can be accounted for by several difficulties in mapping spiral tracers in the Milky Way. In particular, distances to molecular clouds have proven very challenging to determine due to the kinematic distance ambiguity \citep{sanders85, solomon87, clemens88}, and to streaming motions affecting the accuracy of kinematic distances \citep{reid09}. In addition, the lack of sampling, resolution, and sensitivity of CO surveys has made the derivation of the physical properties and structure of molecular clouds difficult \citep{heyer09}. \\
 \indent The lifetimes of molecular clouds also constitute a subject open to debate: are they long-lived  (lifetime $>$ 10$^8$ years) or short-lived (lifetime of 10$^6$ - 10$^7$  years) ? We still do not know whether molecular clouds are transient features or gravitationally bound structures. Simulations by \citet{dobbs06} suggest that molecular clouds are unbound. Only clumps embedded in molecular clouds are gravitationally bound and are subsequently able to form stars. On the other hand, simulations by \citet{tasker09} suggest that molecular clouds are gravitationally bound. The gravitational state of a molecular cloud is described by its virial parameter. The virial parameter essentially represents the ratio of kinetic to gravitational energy in a molecular cloud. If it is greater than 1, the molecular cloud is not gravitationally bound. If it is smaller than 1, the molecular cloud is gravitationally bound. \\
\indent The physical properties of molecular clouds (mass, size, density, temperature, virial parameter...) should not only reflect their formation and dynamical evolution, but also provide key information about the spiral structure of the Milky Way. Theoretical models predict that warm, diffuse gas is compressed in spiral arms, leading to the formation of atomic clouds that give rise to molecular clouds within 10 My \citep{dobbs06, glover10}. The onset of star formation in high density regions created by turbulent density fluctuations and shocks occurs within a few global free fall times \citep{klessen00}.  By the time a molecular cloud has formed and star formation been triggered, the molecular cloud has already reached the leading edge of the spiral arm where it formed, and starts to enter the inter-arm space. In the meantime, massive star formation photo-evaporates, photo-ionizes, and dynamically disrupts the molecular cloud. As a result, molecular clouds should be rapidly (within a few million years) disrupted in the inter-arm space. If these theoretical predictions are correct, molecular clouds located inside spiral arms should be somewhat gravitationally bound while inter-arm molecular clouds should exhibit signs of disruption due to massive star-formation activity and dynamical disruption.\\
\indent  Physical properties of molecular clouds have previously been investigated by \citet{solomon87} using the University of Massachusetts-Stony Brook (UMSB) \COTJ Galactic Plane Survey \citep{sanders86, clemens86}. Being undersampled in solid angle (with a resolution of 48" and a grid spacing of 3'), the UMSB survey did however not yield an accurate derivation of the physical properties of molecular clouds. In addition, molecular clouds detected in the UMSB survey suffered from blending, particularly near the tangent point, due to the use of an optically thick tracer (\COT). The Boston University - Five College Radio Astronomy Observatory (BU-FCRAO) Galactic Ring Survey \citep[GRS;][]{GRS} is the first large scale \COJ survey that is both fully sampled and has an angular resolution sufficient to resolve the small scale structure of molecular clouds (resolution of 0.2 pc at a distance of 1 kpc). With a resolution of 46", a grid spacing of 22", a sensitivity of 0.13 K, and a coverage of 75 square degrees, it  is a reliable data set to probe both the structure of molecular clouds and the Galactic distribution of molecular clouds. In addition, because the GRS uses an optically thin tracer ($\tau$ ($^{13}$CO) $\simeq$ 1), and offers a spectral resolution superior to previous surveys, it does not suffer from blending as much as the UMSB, and can be used to reliably identify molecular clouds. \\
\indent In this paper, we derive the physical properties (mass, radius, surface density, ...) and the Galactic mass distribution of a sample of 750 molecular clouds identified in the GRS \citep{rathborne08}, for which kinematic distances are available from \citet{RD2009}. The paper is organized as follows. The data, methodology and conventions used to derive the physical properties of 580 molecular clouds detected in the GRS and covered by the UMSB survey are described in Sections \ref{data} and \ref{derivation}, and the catalog is shown in Section \ref{results}. In section \ref{fractal}, we discuss the fractal structure the ISM, which can be derived from the relation between their sizes and masses. This relation allows us to derive masses for an additional 170 GRS molecular clouds not covered by the UMSB survey in Section \ref{fractal_masses}. The histograms and range of values for the physical properties of molecular clouds are derived in Section \ref{histos}. In Section \ref{galactic_distribution}, we derive the Galactic surface mass density of molecular gas and how it correlates with different models of the spiral structure of the Milky Way. Section \ref{rg_prop_section} examines the variations of the physical properties of the GRS molecular clouds with Galactocentric radius. Finally, the limitations and caveats of this analysis are discussed in section \ref{discussion_properties}. Section \ref{conclusion} concludes this analysis.

 \section{Molecular line data}\label{data}

\subsection{$^{13}$CO J = 1 $\rightarrow$ 0 GRS Data}
\indent The sample of molecular clouds used in this analysis was identified by their \COJ  emission in the BU-FCRAO GRS \citep{GRS}. The GRS was conducted using the FCRAO 14 m telescope in New Salem, Massachussetts between 1998 and 2005. The survey, which used the SEQUOIA multipixel array, covers the range of Galactic longitude 18\degs $\leq$  $\ell$  $\leq$  55.7\degs and Galactic latitude $-$1\degs $\leq$  $b$  $\leq$  1\degn. The survey achieved a spatial resolution of 46", sampled on a 22'' grid, and a spectral resolution of 0.212 \kms for a noise variance $\sigma$ (T$_A^*$) = 0.13 K ($\sigma_{T_{mb}} = 0.26$ K accounting for the main beam efficiency of 0.48). The survey covers the range of velocity $-$5 to 135 \kms for Galactic longitudes $\ell$ $\leq$ 40\degs and $-$5 to 85 \kms for Galactic longitudes $\ell$ $\geq$  40\degn. \\
\indent Because the GRS uses $^{13}$CO, which has an optical depth 50 times lower than that of \COT, it allows for a better detection and separation of molecular clouds both spatially and spectrally than previous \COT surveys. Using the algorithm CLUMPFIND \citep{williams94} applied to the GRS data smoothed to 0.1\degs spatially and to 0.6 \kms spectrally, 829 molecular clouds were identified by \citet{rathborne08}. CLUMPFIND identifies as molecular clouds a set of contiguous voxels (i.e., ($\ell$, $b$, $v$) positions) with intensity values higher than a given threshold, which has to be determined empirically so as to best identify molecular clouds at all levels of emission. We refer the reader to \citet{rathborne08} for the details of the identification procedure. Molecular cloud parameters such as Galactic longitude, latitude, and velocity of the \CO emission peak were estimated by CLUMPFIND. Individual \CO data cubes extracted from the GRS, and covering only the Galactic longitude, latitude, and velocity range of each individual molecular cloud were also created by \citet{rathborne08}. We use kinematic distances to 750 out of the 829 GRS molecular clouds derived in \citet{RD2009} using the \citet{C85} rotation curve and \his self-absortion to resolve the kinematic distance ambiguity. \\

\subsection{$^{12}$CO J = 1 $\rightarrow$ 0 UMSB data}
\indent We make use of the \COT UMSB survey \citep{sanders86, clemens86}, a joint program between FCRAO and the State University of New York at Stony Brook performed between November 1981 and March 1984. All of the observations were obtained using the FCRAO 14 m telescope. A grid sampled every 3' covering the range 18\degs $<$ $\ell$ $<$ 55\degs and $-$1\degs $<$ b $<$ $+$1\degs  was observed with a velocity resolution of 1 \kms and a spatial resolution of 44". The sensitivity of the observations is 0.4 K per velocity channel. The UMSB survey only covers the velocity range -10 \kms $<$ V$_{LSR}$ $<$ 90 \kmsn, which excludes the GRS molecular clouds located at low longitude and at the tangent point. \\

\section{Derivation of  the physical properties of GRS molecular clouds}\label{derivation}

\subsection{Excitation temperatures}
\indent The \COTJ excitation temperature of our sample of 580 molecular clouds was derived from the \COT brightness temperature, with the assumption that the \COT line is optically thick (i.e $\tau$ ($^{12}$CO) $\gg$ 1). In the optically thick regime, the observed \COT brightness temperature $T_{12}$ and the excitation temperature $T_{ex}$ are related by \citet{rohlfs03}: 
\begin{equation}
T_{ex} = 5.53 \: \frac{1}{ln \left (1+\frac{5.53}{T_{12}+0.837} \right )}
\label{tex_eq}
\end{equation}

\noindent where both temperatures are in $K$. This equation includes a background subtraction accounting for the cosmic microwave background at T = 2.73 K.  The excitation temperature was calculated at each voxel (i.e., ($\ell$, $b$, $v$) position) associated with each molecular cloud in the UMSB data resampled to the GRS grid. The \COT excitation temperature was used as a proxy for the \CO excitation temperature. The \CO and \COT excitation temperature should be identical in local thermal equilibrium since the energy levels of the two isotopomers are roughly the same. In reality, density gradients in molecular clouds and non-local thermal equilibrium situations can lead to differences in the excitation temperatures of \CO and $^{12}$CO. These effects are described in Section \ref{subthermal_section}. The mean excitation temperature of each molecular cloud was computed by averaging the excitation temperature over all the voxels that contain a \CO brightness temperature greater 4$\sigma$ or, accounting for the main beam efficiency, a minimum brightness temperature T$_{13}$ $>$ 1 K. 

\subsection{\CO optical depths}
\indent The \CO optical depth was derived using i) the \COT excitation temperature as a proxy for the \CO excitation temperature, and ii) the GRS \CO data. Since the UMSB and GRS data have nearly the same beam-width, corrections to account for different beam dilution between the two surveys can be neglected. The \CO optical depth is thus given by \citet{rohlfs03}:

{\footnotesize
\begin{equation}
\tau_{13} (\ell, b, v) = - ln \left( 1 -  \frac{0.189 \: T_{13}(\ell, b, v)}{\left (e^{\frac{5.3}{T_{ex}(\ell, b, v)}}  -1 \right )^{-1}-0.16 } \right )
\label{tau_eq}
\end{equation}
}

\noindent where $\tau_{13}$ is the \CO optical depth and T$_{13}$ is the background subtracted \CO brightness temperature in K. The \CO optical depth is evaluated at each GRS voxel ($\ell$, $b$, $v$) associated with each molecular cloud (i.e., in the individual data cubes associated with each molecular cloud). The line center optical depth was computed at each pixel (i.e., position ($\ell$, $b$) on the sky) where the \CO integrated intensity is greater than 4$\sigma$  $=$ 0.23$\sqrt{N_v}$ K \kmsn, where $N_v$ is the number of channel in the cube.  We also computed the average of the line center optical depth over all GRS pixels associated with each cloud where the \CO integrated intensity is greater than 4$\sigma$.

\subsection{\CO column densities}
\indent From the optical depths and excitation temperatures, \CO column densities were derived at each GRS pixel ($\ell$, $b$) associated with each molecular cloud \citep{rohlfs03}:

{\footnotesize
\begin{equation}
\label{column}
\frac{N(^{13}CO) (\ell, b)}{\mathrm{cm}^{-2}} = 2.6 \times 10^{14}\:  \int{\frac{T_{ex} (\ell, b, v) \:\tau_{13}(v)}{1-e^{\frac{-5.3}{T_{ex} (\ell, b, v)}}} \frac{dv}{\mathrm{km} \: \mathrm{s}^{-1}}}
\end{equation}
}

\noindent \CO column densities are computed by integrating Equation \ref{column} only over GRS voxels associated with each cloud where the \CO brightness temperature is greater than 4$\sigma$ $=$ 1 K.

\subsection{Radii of molecular clouds}\label{size_section}
\indent The solid angle subtended by each GRS molecular cloud is computed by counting the number $N_{pix}$ of GRS pixels (i.e., ($\ell$, $b$) positions on the sky) associated with each molecular cloud that contain a \CO integrated intensity greater than 4$\sigma$ $=$ 0.23$\sqrt{N_v}$ K \kmsn:
\begin{equation}
\Omega  = N_{pix} \Delta l \Delta b
\end{equation}
where $\Delta \ell$ and $\Delta b$ are the angular width of the pixels. Knowing the distance $d$ to the molecular clouds, the area $A$ and linear equivalent radius $R$ of the molecular clouds are given by:
\begin{eqnarray}
A = \Omega d^2 \\
R = \sqrt{\frac{A}{\pi}}
\end{eqnarray}

\subsection{Velocity dispersion}\label{vel_disp_section}
The one-dimensional, intensity weighted, velocity dispersion of a molecular cloud is defined as:

\begin{equation}
\sigma_{v_{1D}}^2 = \frac{\sum_{T_{13}(\ell, b, v) > 1 K}{ T_{13}(\ell, b, v) (v - <v>)^2}}{\sum_{T_{13}(\ell, b, v) >1 K} {T_{13}(\ell, b, v)}}
\end{equation}
where $T_{13}$ is the \CO brightness temperature in K, and only positions and velocity channels where the \CO main beam temperature is greater than 4$\sigma$ $=$ 1 K are taken into account. The average radial velocity $<v>$ is given by:

\begin{equation}
<v> = \frac{\sum_{T_{13}(\ell, b, v) > 1 K}{ T_{13}(\ell, b, v) v}}{\sum_{T_{13}(\ell, b, v) >1 K} {T_{13}(\ell, b, v)}}
\end{equation}

\subsection{Masses} \label{mass_section}

\indent Molecular hydrogen (H$_2$) and Helium (He) are the main constituents of molecular clouds. In order to derive molecular cloud masses from CO observations, one must therefore make assumptions about the abundance of CO relative to H$_2$. First, we assume that the abundance of \CO relative to H$_2$ and He is uniform. In reality, the CO abundance declines steeply with decreasing $A_V$ due to photo-dissociation by the Galactic radiation field at A$_v$ $<$ 3 \citep{glover10}. In constrast, self-shielded H$_2$ can exist at $A_V$ as low as 0.2 \citep{wolfire10}. As a result, the \CO/H$_2$ abundance is likely to be lower in the envelopes of molecular clouds than in their denser cores. The effects of abundance variations on the estimation of the molecular cloud masses are discussed in Section \ref{abundance}. Nevertheless, for simplicity, constant ratios n($^{12}$CO)/n($^{13}$CO) = 45 and n($^{12}$CO)/n(H$_2$)  =  8 $\times$ 10$^{-5}$ are assumed \citep{langer90, blake87}. A mean molecular weight of 2.72 accounts for the presence of both H$_2$ and He \citep{allen73, simon01}. Under those assumptions, molecular cloud masses are derived from equation \ref{masseq} below using kinematic distances from \citet{RD2009}.

{\footnotesize
\begin{equation}
\label{masseq}
\frac{M}{M_{\odot}} = 0.27 \: \frac{d^2}{\mathrm{kpc}^2}\: \int_{\ell, b, v} { \frac{T_{ex}(\ell, b, v) \: \tau_{13}(\ell, b, v)}{1-e^{\frac{-5.3}{T_{ex} (\ell, b, v)}}}\frac{ dv}{\mathrm{km} \: \mathrm{s}^{-1}}\: \frac{d\ell}{'} \frac{db}{'} }
\end{equation}
}

\noindent In this equation, the integrand is summed over all ($\ell$, $b$) positions where the \CO integrated intensity is greater than 4$\sigma$. Along those sightlines, only velocity channels where the brightness temperature is greater than 4$\sigma$ $=$ 1 K are included in the integration. The second criterion is necessary due to numerical issues with Equation \ref{tex_eq}, \ref{tau_eq}, and \ref{masseq} when the brightness temperature value is negative. Hence, we do not include contributions from the noise in the integration described by Equation \ref{masseq}. However, some noise peaks in voxels isolated from molecular cloud emission, with values greater than 4$\sigma$, still remain even after filtering the integration with the first criterion. To remedy this problem, we filter the line-of-sights ($\ell$, $b$) along which the integration is perform by applying a threshold in \CO integrated intensity (first criterion).

\subsection{Number density and surface mass density}
\indent The mean number density of particles (H$_2$ and He) in the molecular clouds was estimated assuming spherical symmetry via:

\begin{equation}
\frac{n(\mathrm{H}_2 + \mathrm{He})}{\mathrm{cm}^{-3}}  = 15.1 \times \frac{M}{M_{\odot}} \times \left (\frac{4}{3} \pi \:\frac{R^3}{\mathrm{pc}^3} \right )^{-1}
\end{equation}
The surface mass density $\Sigma$ of the molecular clouds is defined as:

\begin{equation}
\frac{\Sigma}{M_{\odot} \: \mathrm{pc}^{-2}} =\left( \frac {M}{M_{\odot}} \right) \left( \frac {A}{\mathrm{pc}^2}\right)^{-1}
\end{equation}

\noindent where the area $A$ of the molecular clouds is calculated as described in Section \ref{size_section}.

\subsection{Virial mass and virial parameter}
\indent Molecular clouds are supported against gravitational collapse by various mechanisms, such as turbulence, thermal gas pressure, and magnetic fields \citep{heitsch01, klessen00}. Observations show that the linewidths of molecular clouds are much wider than their thermal line-widths. Transonic/supersonic turbulence must therefore be the main source of kinetic energy and support in molecular clouds \citep[e.g.,][]{larson81, williams00}. \\
\indent The virial parameter $\alpha$ of a molecular cloud  is the ratio of its virial mass M$_{vir}$ to its mass. It describes the ratio of internal, supporting energy to its gravitational energy. The virial mass of a molecular cloud is defined as the mass for which a molecular cloud is in virial equilibrium, i.e., when the internal kinetic energy $K$ equals half the gravitational energy $U$ (2$K$ + $U$ $=$ 0). It is given by:

\begin{equation}
M_{vir} =  1.3 \frac{R\: \sigma_v ^2}{G} = 905 \frac{R}{\mathrm{pc}}\: \frac{\sigma_{v_{1D}}^2}{(\mathrm{km} \: \mathrm{s}^{-1})^2} 
\end{equation}

\noindent where $R$ is the equivalent radius of a molecular cloud defined in section \ref{size_section} and $\sigma_{v_{1D}}$ is the one-dimensional velocity dispersion defined in section \ref{vel_disp_section}. Note that the 3D isotropic velocity dispersion, $\sigma_v$, is $\sqrt{3}$ times the 1D velocity dispersion, which is measured in our spectroscopic data. For $M$ $>$ $M_{vir}$ ($\alpha$ $<$ 1), 2$K$ + $U$ $<$ 0 and the molecular cloud is gravitationally bound. For $M$ $<$ $M_{vir}$  ($\alpha$ $>$ 1), 2$K$ + $U$ $>$ 0 and the molecular cloud is not gravitationally bound.  \\
\indent The virial mass is proportional to the linewidth and to the radius of a molecular cloud, with the proportionality constant depending on the number density profile. We have computed the CO number density profile by assuming spherical symmetry and deprojecting the column density with an Abel transform. By fitting a power-law to each density profile, we find that the average slope is $-$1.8. To be consistent with past literature \citep{solomon87}, we use the proportionality constant derive from a density profile of slope $-$2 to compute the virial mass.

\subsection{Physical Properties of the GRS molecular clouds}\label{results}
\indent The  first 25 entries of the derived physical properties of the 580 GRS 
molecular clouds covered by the UMSB are listed in Table \ref{mass_subtable}. 
The complete Table (including molecular clouds not covered by the UMSB 
survey, see Section \ref{fractal_masses}), can be found online. The  first four 
columns indicate the molecular cloud name, Galactic longitude, latitude and LSR 
velocity from the \citet{rathborne08} catalog.  Columns 5 indicates the FWHM 
velocity dispersion of the molecular clouds (derived from section 
\ref{vel_disp_section} via a conversion factor of $\sqrt{8 ln(2)}$ between the 1-
$\sigma$ and FWHM velocity dispersion). Column 6 gives the physical radii as 
defined in section \ref{derivation}. Columns 7 and 8 indicate the masses of the 
molecular clouds and their uncertainty. The derivation of error bars on the mass 
estimates are discussed in detail in Section \ref{errors_section}.  Column 9 
provides the mean number density of the molecular clouds. Columns 10 and 11 
provide the mean excitation temperature and the mean \CO center-of-line 
optical depth. Column 12 and 13 show the mean surface density of each cloud 
and the virial parameter. Column 14 contains a flag "i" indicating that the cloud 
is covered by the UMSB survey.

\section{On the fractal dimension of molecular clouds}\label{fractal}
\begin{figure}
\centering
\includegraphics[height= 6.5cm]{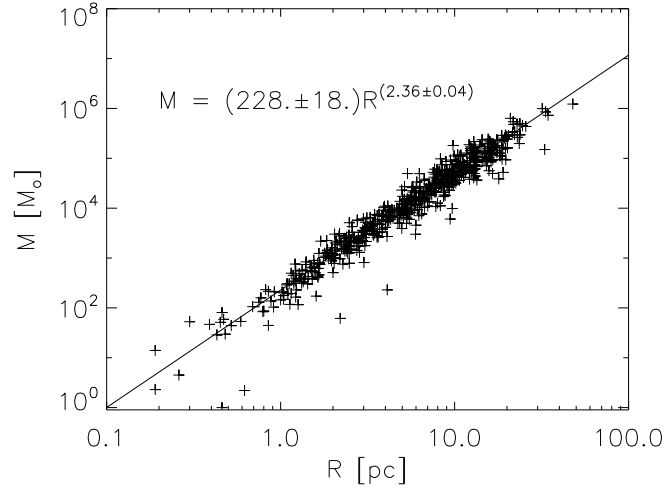}
\caption{Correlation between the masses and radii of molecular clouds. }
\label{mass_siz}
\end{figure}

\indent A dew decades ago, molecular clouds were thought to be isolated, well-defined objects formed by coalescence \citep{oort54, field65}  and sustained in equilibrium by pressure from the hot inter-cloud medium. Since the 1980s, it has been known that molecular clouds are in fact dense sub-structures (n(H$_2$) $>$ a few hundred cm$^{-3}$) in an underlying turbulent, fractal multi gas phase ISM \citep{scalo85, scalo88, falgarone89, scalo90}. Molecular clouds form in gas over-densities resulting from supersonic turbulent flows. When these over-densities reach a visual extinction greater than a few tenths, molecules, such as H$_2$ and CO, become shielded from the photo-dissociating interstellar radiation field by dust \citep{glover10, wolfire10}.  In this respect, molecular cloud boundaries are observed to be fractal \citep{beech87, scalo90, zimmermann92}. The interiors of molecular clouds are also fractal in nature, as shown by their power-law density spectra \citep{brunt10}, energy spectra \citep{rd10}, and size-linewidth relation \citep{larson81, heyer09}. On the smallest scales, very dense (n(H$_2$) $>$ 10$^5$ cm$^{-3}$) molecular cores give birth to star clusters. \\
\indent The fractal dimension of turbulent gas describes how completely it fills space as one zooms down to smaller and smaller scales. In other words, the fractal dimension corresponds to the degree of ``sponginess''. The fractal dimension of molecular clouds has so far been investigated via the perimeter-area relation \citep{beech87, wakker90, bazell88, scalo90, fed09a}, which relates the perimeter of a molecular cloud to its projected area on the sky. It has however been shown by \citet{mandelbrot83} that the radii $R$ and masses $M$ of sub-structures in a fractal are related via $M$ $\propto$ $R^D$. Because molecular clouds are sub-structures in an underlying fractal ISM, one should in principle be able to estimate the fractal dimension of the ISM within the range of spatial scales covered by molecular clouds from the correlation between the radii and masses of molecular clouds. Altough there are, in reality, other non-random mechanisms (e.g., spiral density waves) that modulate the distribution of cloud masses, treating molecular clouds as sub-structures in a fractal is a reasonable 0$^{th}$ order approximation.\\
\indent Figure \ref{mass_siz} shows the correlation between molecular clouds' radii and masses. The radii and masses of GRS molecular clouds are related by a tight power-law correlation: $M$ $=$ 228$\pm$18$R^{2.36\pm0.04}$, of exponent $D=2.36\pm0.04$. Within the error bars, this value is consistent with the value of $D$ derived in the literature \citep{falgarone91, elmegreen96, fed09a}. The slope of the correlation was obtained by applying a chi-square minimization between a linear model and the observed relation between the logarithms of molecular clouds' radii and masses, weighted by the error on the mass. The error on the slope quoted here thus corresponds to the error on the linear fit between $log(L)$ and $log(M)$, including the error on the mass estimation for each molecular cloud. The error cited here is quite small (4\%) compared to the error cited by \citet{elmegreen96} for instance (30\%). The difference is due to the method used to derived $D$. \citet{elmegreen96} derived $D$ from several surveys, some of which do not contain nearly as many molecular clouds as our sample. Hence, the errors quoted for the fractal dimension derived from individual surveys with small samples are larger than our error estimation. In addition, the final value of $D$ quoted by \citet{elmegreen96} corresponds to the average value yielded by all surveys, and the error on this value reflects the dispersion between the different surveys, reduced with different methods and calibrations. However, \citet{elmegreen96} find $D$ $=$ 2.38$\pm$0.09 for galactic clouds from \citet{solomon87} and \citet{dame86}, which is consistent with both our value of $D$ and our error calculation. \\
\indent The value of $D$ between 2 and 3 corresponds to a spongy medium, which fills space more than simple sheets. This seemingly universal value of $D$, also observed in atmospheric clouds,  could result from the intrinsic structure of supersonic, intermittent turbulent flows \citep{sreenivasan86, meneveau89, sreenivasan91}.

\section{Derivation of the masses of molecular clouds outside the UMSB coverage}\label{fractal_masses}
\indent The tight correlation between the radii and masses of molecular clouds derived in Section \ref{fractal} can be used to compute the masses of molecular clouds located outside the UMSB coverage, knowing their radii from Section \ref{size_section}:

\begin{equation}
M = (228\pm18) R^{2.36\pm0.04}
\end{equation}

\begin{figure}
\centering
\includegraphics[width=9cm]{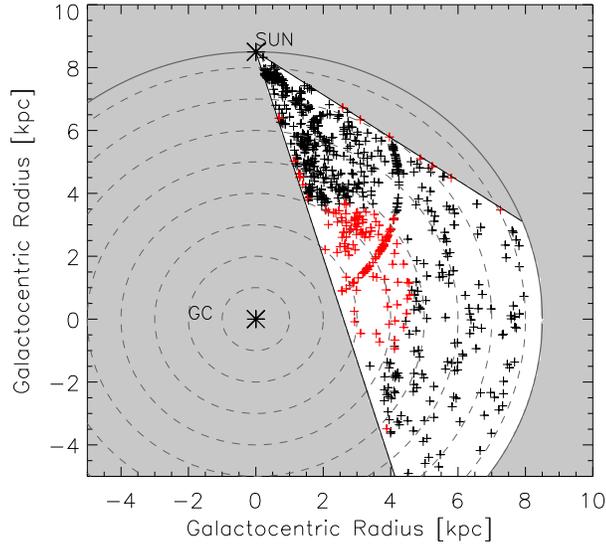}
\caption{Positions of the GRS molecular clouds covered by the UMSB \COT survey (black crosses), and outside of the UMSB coverage (red crosses). The dashed circles indicate galactocentric radii R$_{gal}$ = 1 - 8 kpc, by steps of 1 kpc. The solid circle indicates the solar circle. The white area corresponds to the portion of the Galactic plane covered by the GRS. The artefact produced by the alignment of clouds in an arc of circle is due to the tangent point. Indeed, molecular clouds with radial velocities greater than the radial velocity of the tangent point (due to uncertainties in the rotation curve and non-circular motions) were assigned the distance of the tangent point.  }
\label{plot_umsb_clouds}
\end{figure}

\noindent The density and surface density of those molecular clouds can then be derived using the method described in Section \ref{derivation}. We thus extended our catalog of molecular cloud masses, densities, and surface densities to the 750 objects for which kinematic distances are available from \citet{RD2009}. The online table includes GRS molecular clouds located outside the UMSB survey coverage, which are flagged by an ``o'' in the last column of the online table and have their temperature and optical depth set to zero. Molecular clouds covered by the UMSB are flagged by an ``i''.   Figure \ref{plot_umsb_clouds} shows the spatial distribution of GRS molecular clouds covered by the UMSB survey, and outside its coverage.

\section{Histograms of molecular clouds' physical properties}\label{histos}
\begin{figure}
\centering
\includegraphics[height= 18.5cm]{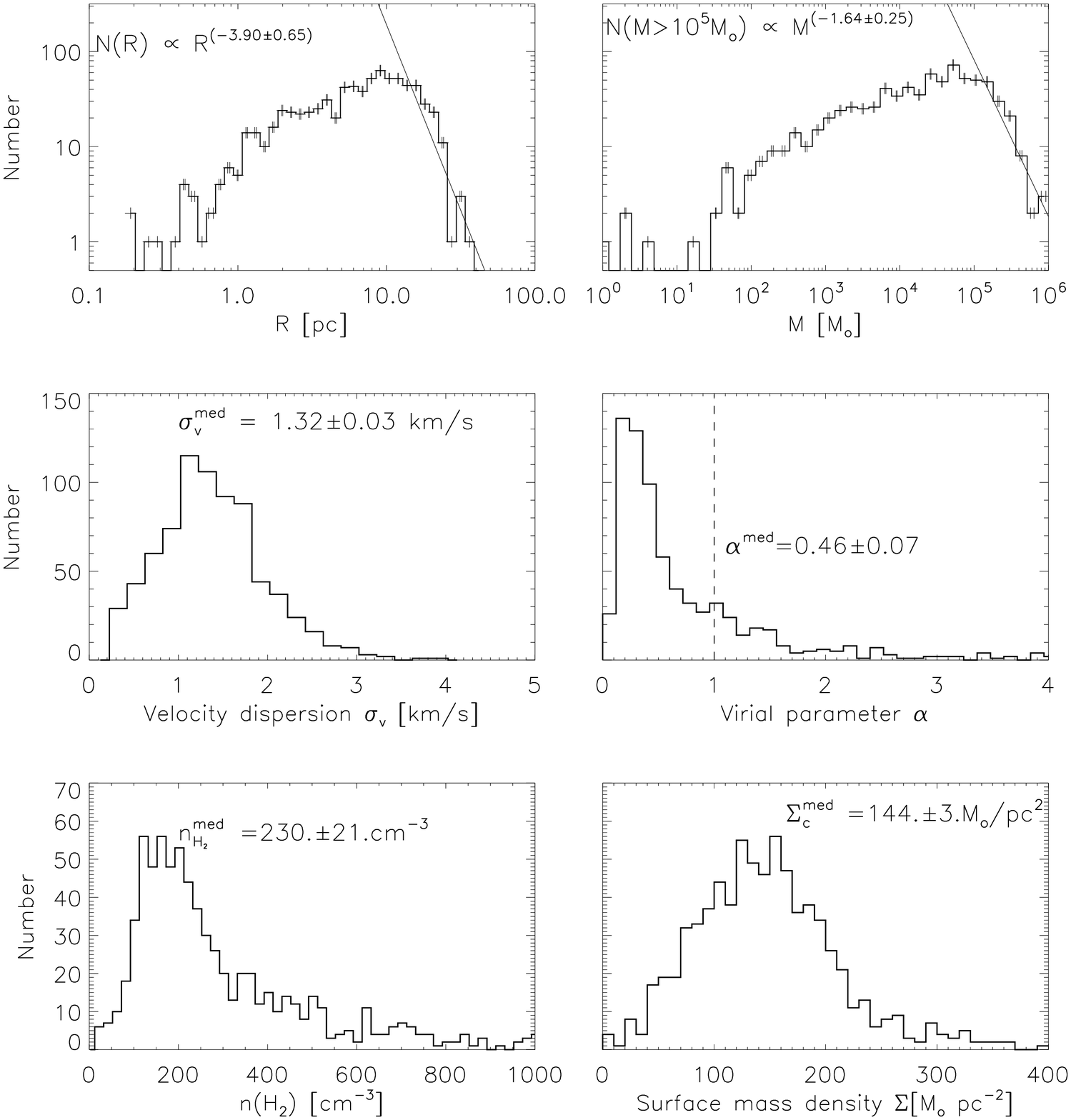}
\caption{Histograms of the physical properties of molecular clouds. In the top two panels, the solid line indicates the best fit to the radius and mass spectra.}
\label{all_histos_distances}
\end{figure}

\begin{figure}
\centering
\includegraphics[width=8cm]{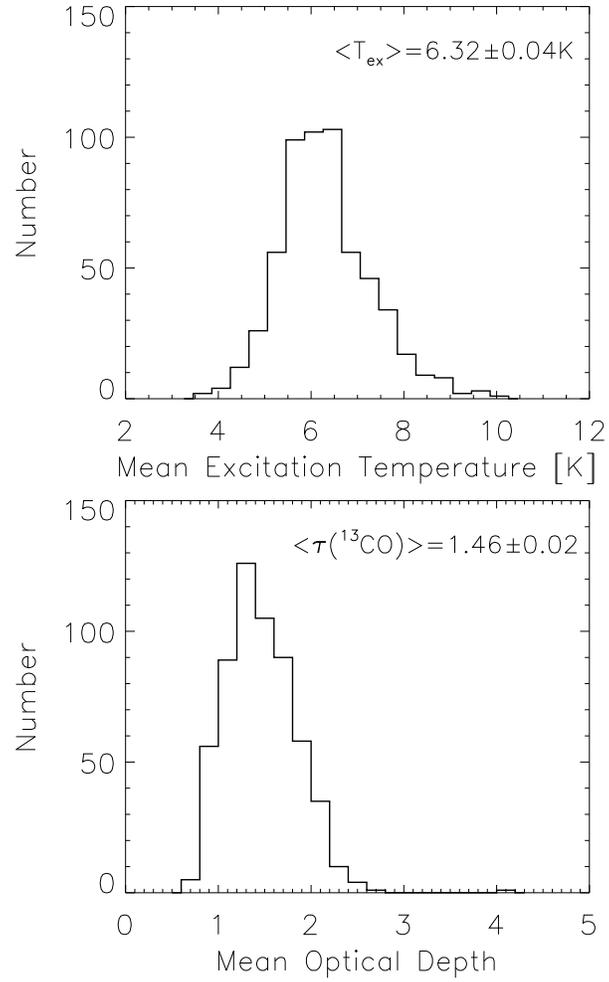}
\caption{Histograms of the excitation temperature (top) and \CO optical depth (bottom) of the molecular clouds averaged over voxels where the brightness temperature is greater than 4$\sigma$ = 1 K.}
\label{histos_excpar}
\end{figure}

\indent This section is intended to provide a range of values for the physical properties of molecular clouds, that may later be used in other studies. For instance, they may be important to constrain models of molecular cloud formation and evolution. The masses and radii of molecular clouds might also be used to predict gamma ray fluxes emanating from the interaction between molecular clouds and cosmic rays \citep{aharonian08, gabici09}. Furthermore, the comparison of the mass spectrum of molecular clouds to that of clumps and cores is essential to understand the fragmentation process that leads from molecular clouds to stars. Last, the radius and mass spectra of molecular clouds have been shown to result from the fractal structure of the ISM \citep{elmegreen96}. They are therefore of great interest if one wants to constrain the fractal dimension of the ISM. The histograms of the physical properties (mass, radius, density, velocity dispersion, surface mass density, and virial parameter) of the GRS molecular clouds are shown in Figures \ref{all_histos_distances} and  \ref{histos_excpar}.

\subsection{Radius and mass distributions}
\indent The top panels of Figure \ref{all_histos_distances} show the radius and mass spectra of the sample of 750 GRS molecular clouds, $\Psi (R)$ = $dR/dln(R)$ and $\Phi (M)$  = $dN/dln(M)$. For M $>$ 10$^5$ \Msun, the mass spectrum follows a power law: $\Phi (M) \propto M^{-1.64 \pm 0.25}$. The radius spectrum also follows a power-law for R $>$ 10 pc: $\Psi (R) \propto R^{-3.90 \pm 0.65}$. Within the error bars, the slope of the mass spectrum derived in this paper is consistent with the value of $-$1.5 obtained in previous work \citep[e.g.,][]{sanders85, solomon87, williams97}. The slope of the radius spectrum is higher than the value obtained by \citet{sanders85}, but is consistent, within the error bars, with the radius distribution obtained by \citet{heyer01}.\\
\indent  Our sample of molecular clouds is complete above the turn-over mass of M$_{to}$ = 4$\times$10$^4$ \Msun, such that the slope of the mass spectrum of molecular clouds should not be affected by a lack of completenesss. Our sample of molecular clouds was identified in a version of the GRS smoothed to 6' spatially and to 0.6 \kms spectrally. GRS molecular clouds were detected by CLUMPFIND as contiguous voxels of brightness temperature greater than 0.2 K \citep{rathborne08}. \citet{rathborne08} furthermore applied the condition that molecular cloud candidates detected by CLUMPFIND must contain at least 16 smoothed voxels in order to be identified as a molecular cloud. Assuming that the \CO line is optically thin, the minimum mass of a molecular cloud is given by:

{\footnotesize
\begin{equation}
\frac{M}{M_{\odot}} \geq  0.05 \: \frac{d^2}{\mathrm{kpc}^2}\: T_{ex} \: e^{\frac{5.3}{T_{ex}}} \times 16 T_{min} \frac{\Delta \ell}{'}\frac{\Delta b}{'}\frac{\Delta V}{\mathrm{km} \mathrm{s}^{-1}} 
\end{equation}
}

\noindent where $d$ is the distance in kpc, $T_{ex}$ is the excitation temperature in K, and $T_{min}$ = 0.2 K is the threshold brightness temperature (corrected for beam efficiency). Hence, with  $T_{ex}$ = 6.32 K  (the average value observed in the GRS), $\Delta \ell$ $=$ $\Delta b$ = 0.1\degn,  and $\Delta V$ = 0.6 \kmsn, the completeness limit is $M_{min}$ $=$ 50 $d_{kpc}^2$, where $d_{kpc}$ is the distance in kpc. Thus, $M_{min}$ $=$ 200 \Msu at 2 kpc, 1250 \Msu at 5 kpc, 5000 \Msu at 10 kpc, and 11250 \Msu at 15 kpc. Since the maximum distance probed by the GRS is 15 kpc (based on the GRS Galactic longitude and velocity coverage), the turn-over mass is greater than the completeness limit of the GRS, and the slope of the mass spectrum of molecular clouds above the turn-over mass should not be affected by completeness effects.\\
\indent This calculation of the completeness limit does not take into account confusion, which in reality has an effect on our data. Confusion is more pronounced near the tangent point, where large physical separations correspond to small radial velocity differences. There is also the problem of molecular clouds that have similar radial velocities, but are located on either side of the tangent point, at the near and far kinematic distances.  In these cases, if the velocity difference between the two clouds is small enough (typically less then the line-width of the clouds), the clump-finding algorithm used to identify the GRS clouds, CLUMPFIND, will blend these two clouds into a single object. The extent of the calculation and modeling required to estimate how confusion affects the completeness limit by far exceeds the scope of this paper. The blending of two molecular clouds depends on many parameters such as the geometry (their location in the Galaxy), the structure of large scale galactic features, the distance and radial velocity of the clouds (related by the rotation curve), their line-widths, radii, angular separation, and the parameters used in CLUMPFIND. Nonetheless, confusion in the GRS, which uses \CO as a tracer, is not nearly as severe as in surveys using the optically thick \COT as a tracer for molecular gas.\\
\indent It is worth mentioning that the GRS has  a limited field-of-view, only covering the Galactic latitude range $-1$\degs $<$ $b$ $<$ 1\degn. This in principle imposes a limit on a maximum cloud's radius, $R_{max}$, and mass, $M_{max}^{FOV}$,  detectable by the GRS. We can use the relation between a cloud's mass and radius derived in Section \ref{fractal} to estimate this maximum mass, with $R_{max}$ $=$ $\pi d/180$.  Thus, we find $M_{max}^{FOV}$ $=$ 1.96$\times10^5d_{kpc}^{2.36}$. We will however show in Section \ref{bias_section} that this limit is never reached, and thus does not affect our sample of molecular clouds and their physical properties. 

\subsection{Virial parameter}
\indent The fourth panel of Figure \ref{all_histos_distances}  shows the histogram of the 750 molecular clouds' virial parameter, the median value of which is 0.46$\pm$0.07. 70\% of our molecular cloud sample (both in mass and number) have virial parameters $<$ 1. This analysis thus suggests that most of the molecular mass contained in identifiable molecular clouds is located in gravitationally bound structures. \\

\subsection{Number density and surface mass density}
\indent The bottom left panel shows the mean density of H$_2$ in our sample of 750 molecular clouds, the median of which is 231 cm$^{-3}$. This value is well below the critical density of the \COJ transition, n$_{cr}$ = 2.7$\times10^{3}$ cm$^{-3}$, suggesting that the gas with density $n$ $>$ n$_{cr}$ is not resolved by a 48'' beam (0.25 pc at $d$ $=$ 1 kpc), and that its filling factor is low.\\
\indent The bottom right panel shows the surface mass density of the molecular clouds, with a median of 144 \Msu pc$^{-2}$. Using the Galactic gas-to-dust ratio $<N_H/A_V>$ $=$ 1.9$\times10^{21}$ cm$^{-2}$ mag$^{-1}$ \citep{whittet03}, this corresponds to a median visual extinction of 7 mag. This value is consistent with the prediction from photo-ionization dominated star formation theory \citep{mckee89}. A median surface mass density of 140 \Msu pc$^{-2}$ is lower than the median value of 206  \Msu pc$^{-2}$ derived by \citet{solomon87} based on the virial masses of a sample of molecular clouds identified in the \COT UMSB survey.  Note that \citet{solomon87} originally found a median surface mass density of 170 \Msu pc$^{-2}$, assuming that the distance from the sun to the Galactic center is 10 kpc. Assuming a Galactocentric radius of 8.5 kpc for the sun, this value becomes 206  \Msu pc$^{-2}$ \citep{heyer09}.  The median surface density derived  here is also higher than the value of 42  \Msu pc$^{-2}$ derived by \citet{heyer09}, who re-examined the masses and surface mass densities of the \citet{solomon87} sample using  the GRS and a method similar method to ours. Similar to our analysis, \citet{heyer09} estimated the excitation temperature from the \COT line emission and derived the mass and surface density from \CO GRS measurements and the excitation temperature. \\
\indent For the \citet{solomon87} molecular cloud sample, \citet{heyer09} found a median surface density of 42 \Msu pc$^{-2}$ using the area A1 (the 1 K isophote of the \COT line) defined by \citet{solomon87} to compute masses and surface mass densities.  However, computing surface mass densities within the half power \COT isophote (A2) yields a median surface mass density close to 200 \Msu pc$^{-2}$ \citep[see Fig. 4 of][]{heyer09}.   It is thus likely that the discrepancy between the surface densities derived here, in \citet{heyer09}, and in \citet{solomon87} be explained by the different methods and thresholds used to compute the molecular clouds' properties.   The \COT line is about 4-5 times as bright as the \CO line \citep{liszt06}. As a result, it is likely that a significant fraction of the area A1 used by \citet{solomon87} to compute the molecular clouds' masses using the \COT line only by assuming that molecular clouds are virialized does not exhibit \CO line emission above the detection threshold of the GRS. Since the \citet{heyer09} derivation of molecular clouds' masses is based on \CO as a tracer of molecular gas column density, this would result in a dilution of the surface mass density derived from \CO, which would appear lower compared to the surface density derived from the brighter \COT emission by \citet{solomon87}. This effects likely contributes to the median surface density in \citet{heyer09} being lower than in \citet{solomon87} and in our analysis. Since our derivation of the molecular clouds' mass, radius, and surface mass density is only performed within the 4$\sigma$ contour of the \CO integrated intensity, it is likely biases towards the most opaque regions of molecular clouds, where the surface mass density is higher than in the envelopes, thus raising the median value of the surface mass density compared to \citet{heyer09}.

\subsection{Excitation temperature and optical depth}

\indent Figure \ref{histos_excpar} shows the histograms of the excitation temperature and \CO optical depth of the 580 GRS molecular clouds covered by the UMSB survey. The mean excitation temperature is 6.32$\pm$0.04 K, and the mean optical depth of the \CO line is 1.46$\pm$0.02. Although the \CO line is less oqaque than its \COT counterpart, optical depth effects should be taken into account in the derivation of \CO column densities from \CO spectral line mapping. \\
\indent \citet{rathborne08} previously derived the excitation temperature and \CO optical depth of the GRS catalog of molecular clouds, and found a lower mean optical depth (0.13), and a higher mean excitation temperature (8.8 K). The difference between the results presented here and in \citet{rathborne08} can be explained by the use of different methods. \citet{rathborne08} first computed the mean \CO and \COT brightness temperatures each GRS molecular cloud, $<T_{13}>$ and $<T_{12}>$ before applying Equations \ref{tex_eq} and \ref{tau_eq} to obtain the ``mean'' excitation temperature and optical depth, $T_{ex}(<T_{12}>)$ and $\tau_{13}(<T_{13}>, <T_{ex}>)$. In contrast, we first computed the excitation temperature $T_{ex}$ and the \CO optical depth $\tau_{13}$ from the \CO and \COT brightness temperatures using Equations \ref{tex_eq} and \ref{tau_eq} {\it at each voxel} associated with a particular molecular cloud. Only then did we average the excitation temperature over voxels with \CO brightness temperature above the noise threshold (4$\sigma$), and the line center optical depth over pixels with \CO integrated intensity above the 4$\sigma$ noise level to obtain $<T_{ex}(T_{12})>$ and $<\tau_{13}(T_{13}, T_{ex})>$. Because Equations \ref{tex_eq} and \ref{tau_eq} are non linear with $T_{12}$, $T_{13}$ and $T_{ex}$, the difference between performing the average before or after applying the equation can be quite large. For instance, for GRSMC G053.59+00.04, we find  $<\tau_{13}(T_{13}, T_{ex})>$ $=$ 1.78, while  $\tau_{13}(<T_{13}>, <T_{ex}>)$ $=$ 0.24, close to the typical optical depth found by \citet{rathborne08}.  In this respect, molecular cloud masses recently derived by \citet{urquhart10} likely under-estimate the masses of GRS molecular clouds, since they use the (uniform) excitation temperatures and optical depths derived by \citet{rathborne08}. We find that the masses derived by \citet{urquhart10} are lower than masses derived here by a factor 2 to 3.

\section{Galactic mass distribution of molecular clouds}\label{galactic_distribution}

\begin{figure}
\centering
\includegraphics[width= 9cm]{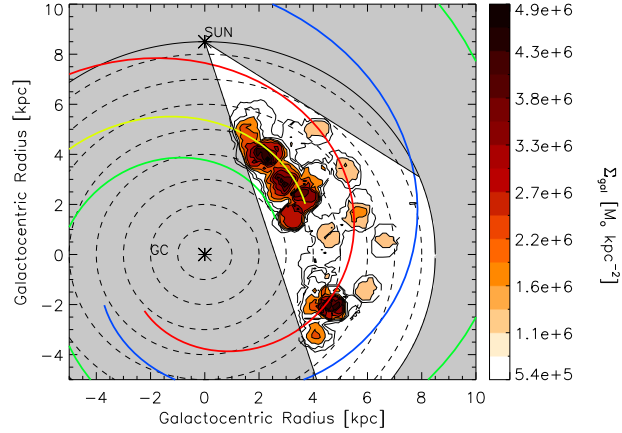}
\caption{Galactic surface mass density of molecular clouds. The contour levels are 10, 20, 30, 40, 50, 60, 70, 80, 90$\%$ of the maximum density (5.3 $\times$ 10$^6$). The dashed lines indicate galactocentric radii of 1, 2, 3, 4, 5, 6, 7, and 8 kpc. The solid line indicates the solar circle. The white area corresponds to the portion of the Galactic plane covered by the GRS. The green, yellow, red, and blue lines represent the 3 kpc arm, the Scutum-Crux arm, the Sagittarius arm, and the Perseus arm from the four-arm model by \citet{vallee95} respectively.  }
\label{mass_clemens_nobg}
\end{figure}

\indent \citet{RD2009}  derived the Galactic \CO surface brightness, which roughly represents the Galactic surface density of molecular gas traced by CO. This analysis, however, did not account for excitation temperature variations and \CO optical depth effects. Due to variations in the heating rate with local environment, the excitation temperature is however likely to vary significantly with Galactocentric radius and local environment. The densest regions of molecular clouds exhibiting \CO optical depths $>$ 1, non-linearities between \CO surface brightness and molecular gas masses should be taken into account. Molecular cloud masses derived in this paper therefore allow for a more accurate and rigorous derivation of the Galactic mass distribution of molecular clouds. We derived the surface mass density of molecular gas in the Milky Way by summing the masses of GRS molecular clouds over circular bins of radius 0.5 kpc, with a sampling  of 0.12 kpc. The resulting galactic surface mass density of molecular gas is shown in Figure \ref{mass_clemens_nobg}. The green, yellow, red, and blue lines represent the 3 kpc arm, the Scutum-Crux arm, the Sagittarius arm, and the Perseus arm from the four-arm model by \citet{vallee95} respectively, which is based on a compilation of different tracers from the literature (e.g., CO, \hiis regions, magnetic fields, electron density etc...).  The locations of the Scutum-Crux and Perseus arms from \citet{vallee95} are coincident with a two-arm model from \citet{benjamin05, benjamin09, churchwell09} based on K-giants and M-Giants star counts from the Galactic Legacy Infrared Mid-Plane Survey Extraordinaire (GLIMPSE). The Sagittarius arm is not detected as an overdensity in the old stellar population, perhaps because the Sagittarius arm is only a gas compression \citep{churchwell09}.  \\
\indent The surface mass density of molecular gas deduced from the GRS appears to be enhanced along the Scutum-Crux and Sagittarius arms. This suggests that, despite the uncertainty in kinematic distances due to non-circular motions near spiral arms and the uncertainty in the rotation curve, molecular clouds are good tracers of the Scutum and Sagittarius arms. Although the Perseus arm has previously been detected in several tracers \citep[e.g., water masers, molecular gas, FIR observations, star counts, see ][]{heyer98, reid09, churchwell09}, the Perseus arm is not detected as a strong enhancement in the Galactic surface density of molecular gas. This could be due to several effects. First, molecular clouds located at 10-15 kpc, on the far side of the Galaxy, tend to have smaller angular sizes than molecular clouds located at closer distances. Because the velocities of near and far molecular clouds located at the same galactocentric radius are the same, it is possible that far molecular clouds be blended together with near molecular clouds and assigned to the near kinematic distance. Second, non-circular motions near the Perseus arm may cause the distance estimate to molecular clouds located in the Perseus arm to be inaccurate. Indeed, there are two large molecular complexes located between the Sagittarius and Perseus arms, which could be associated with the Perseus arm if their distance is systematically under-estimated. Third, the completeness limit of the GRS at 10-14 kpc is 5$\times10^3$-10$^4$ \Msun, such that only a handful of very massive molecular clouds can be detected at the distance of the Perseus arm. Last, the molecular content of the Milky Way in the Galactocentric radius range of the Perseus arm ($R_{gal}$ $=$ 7-8 kpc) is of order $\Sigma_{gal}$ $=$ 1-2 \Msu kpc$^{-2}$, much lower (by at least an order of magnitude, see Section \ref{mass_rg_section}), than in the 5 kpc molecular ring. The combination of these last two effects --- completeness effects and low molecular surface density at $R_{gal}$ $=$ 7-8 kpc --- can be further quantified. Assuming a power-law mass spectrum of exponent $-$1.6 \citep[see Section \ref{histos} and also][]{sanders85, solomon87, williams97},  a Galactic surface density of  molecular gas $\Sigma_{gal}$ $=$ 1-2 \Msu kpc$^{-2}$,  a completeness limit of 10$^4$ \Msu at a distance 10-15 kpc, and a maximum mass of 10$^6$ \Msu (the maximum mass detected in the GRS), we predict that there should be 4 molecular clouds per kpc$^2$ above the completeness limit in the region of the Perseus arm. Indeed, this number would not be sufficient to resolve the Perseus arm. This prediction is also supported by the actual number surface density of molecular clouds observed in the GRS in the Perseus arm region. The fact that the measured surface number density of molecular clouds in the vicinity of the Perseus arm agrees with our simple prediction based on the completeness limit and the azimuthally averaged Galactic surface density of molecular gas suggests that completeness effects and the low Galactic molecular content at Galactocentric radii 7-8 kpc likely account for the non-detection of the Perseus arm in the GRS. However, we do find localized egions with 0-1 cloud kpc$^{-2}$.  Combined with localized enhancements of the number and surface mass density in the inter-arm space between the Sagittarius and Perseus arms, this could suggest that distance uncertainties due to non-circular motion near spiral arms play a role in the non-detection of the Perseus arm. \\
\indent The theoretical implications of the enhancement of the Galactic surface mass density of molecular gas along spiral arms have been discussed in \citet{RD2009}: the confinement of molecular clouds to spiral arms suggests that molecular clouds must form in spiral arms via a combination of hydrodynamical processes due to the compression in the spiral arm, self-gravity, Parker instability, and orbit crowding, and be disrupted in the inter-arm space.

\section{Variations of the physical properties of molecular clouds with Galactocentric radius}\label{rg_prop_section}

\subsection{CO excitation temperature versus Galactocentric radius: effects of star-formation ?}\label{rg_t_section}
\subsubsection{Decline in CO excitation temperature with Galactocentric radius}

\begin{figure}
\centering
\includegraphics[width=8cm]{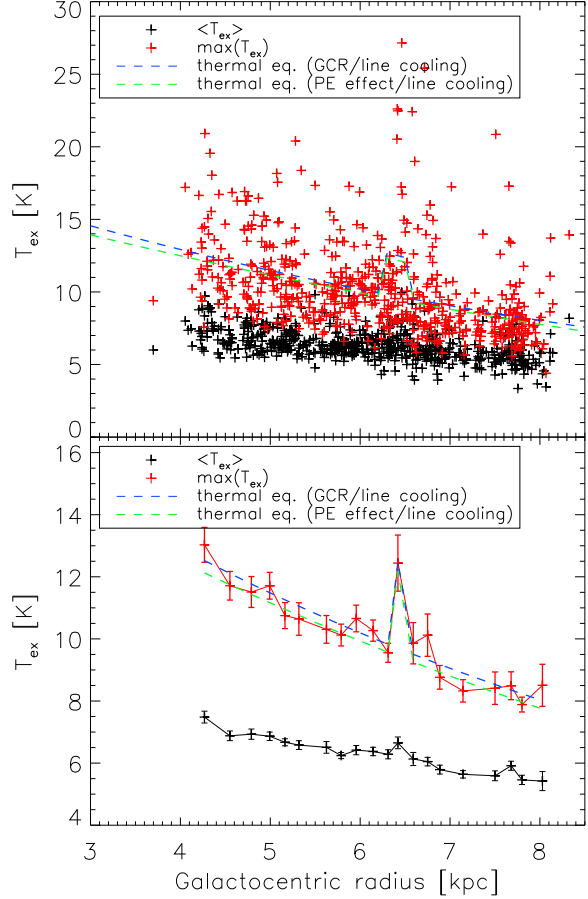}
\caption{Excitation temperature of molecular clouds versus galactocentric radius. The red crosses represent the maximum excitation temperature of each molecular cloud. The black crosses represent the mean excitation temperature of each molecular cloud, averaged over voxels with a brightness temperature greater than 4$\sigma$ = 1 K. In the lower panel, the excitation temperature of GRS molecular clouds were averaged over 0.3 kpc galactocentric radius bins. The CO excitation temperature can be lower than the gas temperature if the local number density is lower than the critical density. The maximum excitation temperature in the clouds shoud be close to the gas temperature. The dashed blue and green lines represent the gas temperature predicted from thermal equilibrium between cooling by CO rotational emission and heating by Galactic cosmic rays (blue) and photo-electric effects from dust grains (green) respectively. Enhancements of 50\% in the GCR flux  and the ISRF were included in the blue and green curves respectively at R$_{gal}$ $=$ 6.4 kpc to explain the observed enhancement in the CO excitation temperature.  }
\label{plot_rg_excpar}
\end{figure}

\indent The equilibrium temperature of molecular clouds results from a balance between heating and cooling.  The heating of gas in molecular clouds is due to: i) electrons ejected from dust grains illuminated by the Interstellar Radiation Field (ISRF) due to the photoeletric effect \citep{bakes94, wolfire03}, ii) H$_2$ photo-dissociation \citep{black77},  iii)  collisions between Galactic Cosmic Rays (GCRs) and molecules \citep{goldsmith78}, iv) UV pumping of H$_2$ \citep{burton90}, and v) formation of H$_2$ on dust grains \citep{hollenbach89}. In the densest, most shielded regions of molecular clouds that are not penetrated by the ISRF, heating by GCRs dominates. In the regions closer to the CO boundary, photoelectric effect likely plays an important role in the heating of the gas. \\
\indent  Fine structure lines of metals, most importantly [\cii] at 158 $\mu$m, and radiative  CO rotational transitions, are responsible for the cooling of gas in molecular clouds.  The decrease in the abundance of metals away from the Galactic center effectively decreases the cooling rate in molecular clouds as their galactocentric radius increases \citep{quireza06}.   \\
\indent Because the gas temperature is the result of the equilibrium between heating and cooling, the effects of the variations in the heating and cooling rates with galactocentric radius on the gas temperature are reflected in the variations of the gas temperature throughout the Galactic plane. The CO excitation temperature however depends on both the gas temperature and the local number density. $T_{ex}$ can be significanlty lower than the gas temperature if the density is lower than the CO critical density, in which case there are not enough particules to collisionally excite the CO molecule to the J=1 level. This effect is known as sub-thermal excitation.  Nonetheless, the CO excitation temperature is likely close to the gas temperature in the dense regions of molecular clouds, which dominate the emission. Hence, the variations of the CO exciation temperature with Galactocentric radius should constrain the balance between gas heating and cooling.\\
\indent Figure \ref{plot_rg_excpar} shows the CO excitation temperature versus $R_{gal}$. Only molecular clouds covered by the UMSB survey, for which we could derive a CO excitation temperature and optical depth, are taken into account. The top panel shows the maximum and mean excitation temperature in each molecular cloud. Since the mean excitation temperature is likely lower than the gas temperature due to sub-thermal excitation effects, the maximum excitation temperature in each cloud, corresponding to the densest regions, probably reflects the gas temperature more accurately. The bottom panel shows these temperatures averaged over 0.3 kpc galactocentric radius bins. In both cases, the CO excitation temperature decreases smoothly with galactocentric radius, from 14 K at $R_{gal}$ $\simeq$ 4 kpc down to 8 K at $R_{gal}$ $\simeq$ 8 kpc (for the maximum excitation temperature). \\
\indent This large-scale decline in the gas temperature away from the Galactic center indicates that the slight decline in metallicity (and hence cooling rate) away from the Galactic center is not large enough to overcome the decrease in the heating rate (due to a decrease in the strength of the ISRF and/or the GCR flux) with Galactocentric radius. We have modeled and quantified the contribution of variations in the GCR flux and the ISRF to the variations of the gas temperature with Galactocentric radius. The variations of the GCR flux have been investigated by \citet{bloemen86} and \citet{strong88}. While the flux of cosmic ray nuclei $\phi_n$ appears constant with Galactocentric radius, the flux of cosmic ray electrons, $\phi_e$, shows a gradient best described by $\phi_e (R_{gal})$ $=$  $\phi_e (R_{\odot})e^{-0.19(R-R_{\odot})}$. Note that \citet{bloemen86} originally cite a gradient of $-$0.16 kpc$^{-1}$, assuming  $R_{\odot}$ $=$ 10 kpc. With the more recent value of $R_{\odot}$ $=$ 8.5 kpc \citep{kerr86}, this corresponds to a gradient of $-$0.19 kpc$^{-1}$.  The heating rate by GCR is then $\Gamma_{GCR}$ $=$ $\phi_e(R_{gal})$ $\Delta_{cr}$ $n(H_2)$ \citep{goldsmith78}, where $\Delta_{cr}$ $=$ 17 - 26 eV is the energy deposited as heat as a result of the ionization by GCRs. The GCR flux at $R_{gal}$ $=$ $R_{\odot}$ is in the range 1.5 - 3 $\times$10$^{-17}$ s$^{-1}$ \citep{goldsmith78}. Hence, the heating rate by GCRs is given by:

\begin{equation}
\Gamma_{GCR} (R_{gal}) = 5\times10^{-28} n(H_2) e^{-0.19(R-R_{\odot})} \: \mathrm{[ergs} \: \mathrm{cm}^{-3} s^{-1} \mathrm{]}
\label{gcr_heating}
\end{equation}

\noindent where we chose a value of $\phi_e \Delta_{cr}$ $=$ 5$\times10^{-28}$ ergs s$^{-1}$, which best describes the excitation temperature trend in our data, and is in the range cited by \citet{goldsmith78}.  \\
\indent The heating rate by photo-electric effect is given in \citet{bakes94}. For neutral grains, the heating rate is given by $\Gamma_{pe}$ $=$ 4.86$\times10^{-26} G_0 (R_{gal}) n_H$ [ergs cm$^{-3}$ s$^{-1}$], in an unattenuated medium, where $G_0$ is the strength of the ISRF in Habing units ($G_0$ varies with Galactocentric radius), and $n_H$ is the hydrogen density. In a molecular cloud of visual extinction A$_V$, the heating rate due to photo-eletric effect is therefore:

\begin{equation}
 \Gamma_{pe} = 4.86\times10^{-26} G_0 (R_{gal}) e^{-5.4 A_V} n_H \: \mathrm{[ergs} \: \mathrm{cm}^{-3} s^{-1}\mathrm{]}
\label{heating_pe}
\end{equation}

\noindent In this equation, we assumed that CO was present for A$_V$ $>$ 1 \citep{wolfire10}, and used the standard Milky Way extinction curve to relate A$_{1000}$, the FUV extinction at 1000 \AA,  and A$_V$ \citep[A$_{1000}$ = 4-5 A$_V$, see][]{gordon03}. The strength of the ISRF, $G_0$, can be observed via the dust temperature \citep{bernard08}: $G_0 \propto T_d^{4+\beta}$, where $\beta$ is the emissivity index of the dust \citep[$\beta$ = 1.5-2, see][]{boulanger96, gordon10}, and $T_d$ is the equilibrium temperature of large dust grains.  In addition, \citet{sodroski97} derived the dust equilibrium temperature versus Galactocentric radius. They found $T_d(R_{gal})$ $=$ 28 K - 6.8 K $\times$ ($R_{gal}$/$R_{\odot}$). Hence, with the local IRSF having strength $G_0(R_{\odot})$ $=$ 1.7, the variations of $G_0$ throughout the Galactic plane are given by:

\begin{equation}
G_0(R_{gal}) = G_0(R_{\odot}) \left (  1.32    - 0.32 \frac{R_{gal}}{R_{\odot}}\right)^{(4+\beta)}
\label{temp_grad}
\end{equation}

 \noindent This expression for $G_0(R_{gal})$ can then be used in Equation \ref{heating_pe} to calculate the heating due to photo-electric effect as a function of $R_{gal}$. \\
\indent The cooling rate by a variety of fine structure atomic and molecular lines is given in \citet{goldsmith78}. For n(H$_2$) $=$ 300 cm$^{-3}$, which is close to the median density ($n_{H_2}^{med}$ $=$ 230 cm$^{-3}$) in our sample of molecular clouds, the cooling rate is $\Lambda$ $=$ 4.7$\times10^{-27} T^{1.6}$ [ergs cm$^{-3}$ s$^{-1}$].\\
\indent In thermal equilbrium, the gas temperature can then be determined by equating the cooling and heating rates: $\Lambda = \Gamma$.  The gas temperature resulting from the balance between heating by GCR and line emission cooling is shown as a blue line in Figure \ref{plot_rg_excpar}. For the H$_2$ number density, n(H$_2$), we use the median value of our sample ($n_{H_2}^{med}$ $=$ 230 cm$^{-3}$) in equation \ref{gcr_heating}. Our simple calculation reproduces the variations of the CO excitation temperature with Galactocentric radius within the error bars. In Figure \ref{plot_rg_excpar}, we also show the gas temperature obtained from the equilibrium between cooling and heating by photo-electric effect (dashed green line). We used Equations \ref{heating_pe} and \ref{temp_grad}, with the median number density found in our molecular cloud sample, and $\beta$ = 1.5 \citep{gordon10}.  In this case, our prediction also reproduces the temperature trend within the error bars. \\
\indent Thus, both a decline in the GCR flux and in the  strength of the ISRF can explain the decrease in CO excitation temperature derived in this analysis based on the GRS.  It is likely that the GCR flux and the strength of the ISRF are correlated, because they both depend on the star formation rate. In this case, the decline in gas temperature with $R_{gal}$ would reflect a decrease in the star formation rate with Galactocentric radius. Indeed, young massive stars produce most of the ionizing and dissociating radiation responsible for gas heating in the envelopes of molecular clouds.  GCRs are thought to be produced in supernova remnants via Fermi acceleration. Since supernovae remnants are the products of massive star-formation, the GCR flux is therefore tied to massive star-formation. Because of the large mean free path of GCRs, however, it is not clear how the heating rate due to GCRs in molecular clouds relates to the local star formation rate. Nonetheless, a gradient of $-$30\%/$R_{\odot}$ in the star formation rate (see Equation \ref{temp_grad}) would explain the decline of the gas temperature with Galactocentric radius.  \\

\subsubsection{Enhancement at $R_{gal}$ $=$ 6.4 kpc}

\begin{figure}
\centering
\includegraphics[width = 9cm]{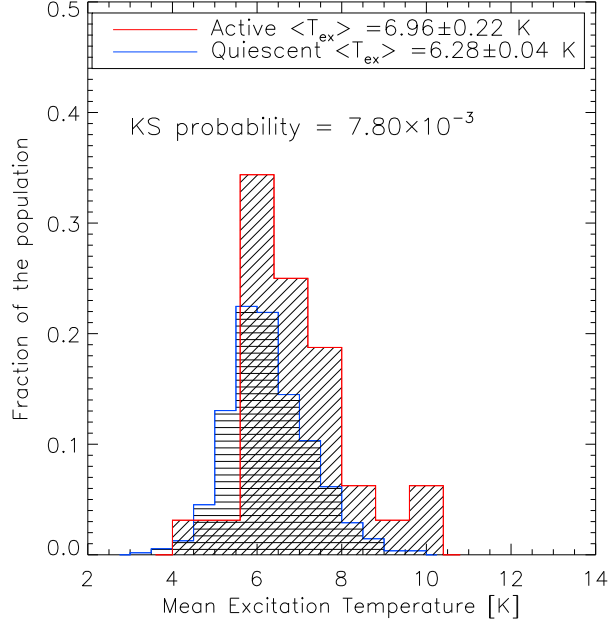}
\caption{Distributions of excitation temperatures of ``active'' (i.e., containing an \hiis region) and ``quiescent'' molecular clouds in red and blue respectively. On average,  active molecular clouds have slighlty higher temperatures.}
\label{compare_tex_hii}
\end{figure}

\indent The maximum CO excitation temperature in each cloud is enhanced by about 2 K at a galactocentric radius of 6.4 kpc. This enhancement is significant at the 3$\sigma$ level with respect to the uncertainty plotted in Figure \ref{plot_rg_excpar}. Over the longitude coverage of the GRS, a galactocentric radius of 6.4 kpc corresponds to the Sagittarius arm. An elevated gas temperature at 6.4 kpc may be related to massive star-formation occuring in the Sagittarius arm (both through the strength of the radiation field and the GCR flux). In Figure \ref{plot_rg_excpar}, an increase of 50\% in the GCR flux (blue line) and in the strength of the ISRF (green line) at $R_{gal}$ $=$ 6.4 kpc reproduces the trend observed in the GRS. \\
\indent The idea that star formation locally increases the gas temperature inside molecular clouds can be further tested by comparing different data sets. \citet{anderson09} found an enhancement of \hiis regions at a galactocentric radius of $\simeq$ 6 kpc. In addition, the 15 hottest molecular clouds located at galactocentric radii 6 - 7 kpc (which stand out in Figure \ref{plot_rg_excpar}) all have confirmed star formation activity - PAH emission in the 8 $\mu$m band, and 24 $\mu$m emission tracing warm dust- in the GLIMPSE and MIPSGAL infrared surveys. Eight of these fifteen molecular clouds are associated with the star formation region W51 located at a galactic longitude of 49\degs - 50\degs and at a distance of 5.5 kpc, which places it near the Sagittarius arm. The seven others belong to different star formation regions and contain known masers and \hiis regions. To test the hypothesis that star formation activity increases the gas temperature of molecular clouds, we compare the excitation temperatures of ``active'' molecular clouds, containing \hiis regions, to the temperature of ``quiescent'' molecular clouds. In \citet{RD2009}, a catalog of such ``active'' and ``quiescent'' molecular clouds was established based on the coincidence between the morphology of the  21 cm continuum and \CO emission, and the velocity of \CO and recombination lines from \citet{anderson09}. This catalog can readily be used to compare the excitation temperatures of active and quiescent molecular clouds. The histograms of the CO excitation temperature of both samples are shown in Figure \ref{compare_tex_hii}. On average, ``active'' molecular clouds have a slightly higher temperature, $<T_{ex}>$ = 6.96$\pm$0.22, compared to ``quiescent'' molecular clouds ($<T_{ex}>$ = 6.28$\pm$0.04). A Kolmogorov Smirnov  test shows that the active and quiescent molecular cloud populations have a less than 1\% chance of being from the same distribution. This difference of temperature between active and quiescent molecular clouds, significant at the 3$\sigma$ level, supports the idea that star formation increases the gas temperature (and hence CO excitation temperature) of molecular clouds.

\subsection{Mass and surface density}\label{mass_rg_section}
\begin{figure}
\centering
\includegraphics[width = \textwidth]{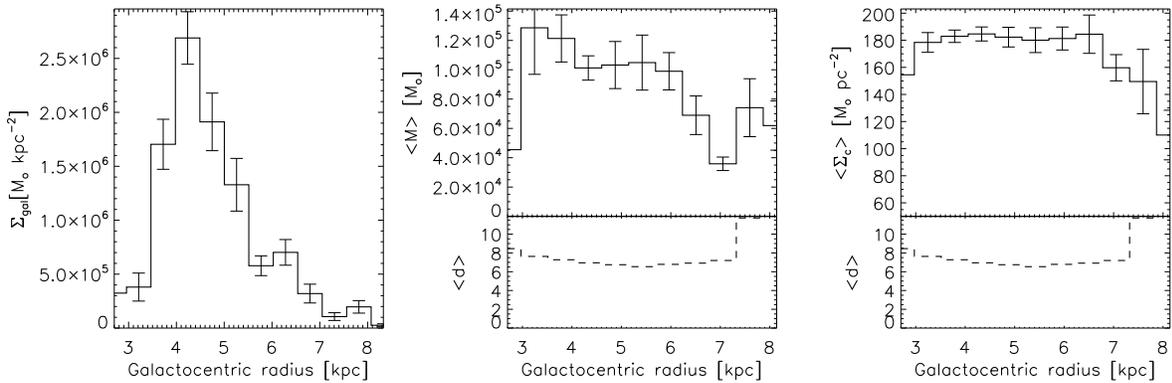}
\caption{Radial variations of i) the Galactic surface mass density of molecular clouds (left), ii) of the molecular cloud masses (middle), and iii) of the molecular clouds' surface mass density (right). The dashed line represents the average $<d>_{0.5 kpc}$ in each 0.5 kpc bin. $<d>_{0.5 kpc}$ is an indicator of how much the average physical properties of molecular clouds within each radial bin are affected by biases (see section \ref{bias_section}). }
\label{plot_rg_masses}
\end{figure}

\indent The variations of the masses and surface mass densities of the GRS molecular clouds as a function of galactocentric radius should help understand and predict the variations in the star formation activity resulting from this reservoir of molecular gas throughout the Galactic disk. The azimuthally averaged Galactic surface mass density of molecular gas, $\Sigma_{\mathrm{gal}}$, was obtained by summing the molecular clouds' masses over 0.5 kpc galactocentric radius bins. Since the range of longitudes covered by the GRS data is limited, each radial bin covers only a limited range in azimuth, which varies with Galacotcentric radius.  To convert the mass measurements to a more uniform measure of gas content, the total mass contained in a radial bin was therefore divided by the surface area covered by the GRS data within the radial bin. As a result, we obtained an azimuthally averaged surface mass density of molecular gas encompassed by GRS molecular clouds at each galactocentric radius.The left panel of Figure \ref{plot_rg_masses} shows $\Sigma_{\mathrm{gal}}$ as a function of galactocentric radius $R_{\mathrm{gal}}$. $\Sigma_{\mathrm{gal}}$ peaks at $R_{gal}$  = 4.5 kpc ($\Sigma_{\mathrm{gal}}$ $=$ 2.5$\times$10$^6$ \Msun kpc$^{-2}$), and decreases steeply with galactocentric radius, down to 10$^5$ \Msun kpc$^{-2}$ at $R_{gal}$ = 7.5 kpc. \\
\indent The middle panel of Figure \ref{plot_rg_masses} shows the molecular clouds' masses averaged over 0.5 kpc galactocentric radius bins. Since each Galactocentric radius bin covers a range of distances, and since the completeness limit varies with distance, only molecular clouds with masses greater than the completeness limit at 15 kpc, or 1.1$\times10^4$ \Msu pc$^{-2}$, were taken into account. The mean molecular cloud mass in each radial bin appears rather constant with $R_{\mathrm{gal}}$ out to 5.5 kpc, at which point the mean molecular cloud mass starts to drop from 10$^5$ \Msu at $R_{\mathrm{gal}}$ = 5.5 kpc down to 5$\times 10^4$ \Msu at $R_{\mathrm{gal}}$ = 7.5 kpc.\\
\indent The right panel of Figure \ref{plot_rg_masses} shows the molecular clouds' surface mass density, $\Sigma_c$, averaged over 0.5 kpc galactocentric radius bins. $\Sigma_c$ increases from 3 kpc to 4 kpc, remains constant between 4 kpc and 6 kpc, and decreases beyond this point,  from 170 \Msu pc$^{-2}$ down to 120 \Msu pc$^{-2}$ at $R_{\mathrm{gal}}$ = 7.5 kpc. It has been suggested by \citep{mckee89} that the star formation rate in a molecular cloud is governed by ambipolar diffusion, and subsequently by the surface mass density. In a molecular cloud with low surface mass density, the ionized fraction sustained by the ISRF penetrating the molecular cloud is high enough to prevent collapse and subsequent star formation due to magnetic support. In contrast, star formation proceeds more rapidly in molecular clouds that exhibit a high surface mass density. The decline in the average molecular cloud surface mass density with galactocentric radius beyond 6 kpc may therefore suggest a decrease in the star formation rate with Galactocentric radius. \\

\section {Discussion}\label{discussion_properties}

\indent The purpose of this section is to emphasize the major assumptions and limitations of the results presented in this paper. Observational biases affecting the physical properties of molecular clouds located at different distances, as well as fundamental assumptions and uncertainties made in the estimation of the physical properties of molecular clouds are discussed. 

\subsection{Biases}\label{bias_section}

\indent The top panel of Figure \ref{bias1} shows the mass versus the distance to each molecular cloud. The mass (and radius) of the GRS molecular clouds is observed to on average increase with distance. This is due to i) a Malmquist bias resulting from the increasing completeness limit and probability of observing a massive cloud with distance, and to a minor extent ii) the effects of finite resolution and the method used to identify molecular clouds. Indeed, it is hard to define one set of parameters for CLUMPFIND (spatial and spectral smoothing, brightness threshold and brightness increment) that identify a molecular cloud at all emission levels.

\subsubsection{Modeling biases}
\indent In order to analytically describe the effects of increasing completeness limit and probability of observing rarer, more massive clouds as a function of distance, we assume a power-law mass spectrum, $\phi (M)$ = $\phi_0 M^{-1.64}$ (see Section \ref{histos}). The average mass of a molecular cloud detected by the GRS at a distance $d$ in the GRS field is:

\begin {equation}
<M>(d) = \frac{\int_{M_{min}(d)}^{M_{max}(d)} {M\phi(M) dM}}{\int_{M_{min}(d)}^{M_{max}(d)} {\phi(M) dM}}
\label{average_mass_eq1}
\end{equation}

\noindent or

\begin {equation}
<M>(d)  =\frac{M_{max}(d)^{0.36}-M_{min}(d)^{0.36}}{M_{min}(d)^{-0.36} - M_{max}(d)^{-0.36}}
\label{average_mass_eq2}
\end{equation}

\noindent where $M_{min}$(d) is the minimum mass that can be detected by the GRS at distance $d$ i.e., the completeness limit. $M_{min}(d)$ is given by $M_{min}(d)$ $=$ 50$d_{kpc}^2$ (see Section \ref{histos}), where $d_{kpc}$ is the distance in kpc. $M_{max}$ is a physical upper limit on the masses of molecular clouds used to integrate Equation \ref{average_mass_eq1}, which would diverge for $M_{max}$ $=$ $\infty$. Indeed, molecular clouds cannot grow infinitely due to disruptive mechanisms (e.g., turbulence, Galactic shear), the limited molecular scale height and limited molecular content of the Milky Way. In addition, the field-of-view of the GRS is limited to galactic latitudes $-$1\degs $<$ $b$ $<$ 1\degn, such that the maximum mass detectable in the survey cannot exceed the limit derived in Section \ref{histos}: $M_{max}^{FOV}$ $=$ 1.96$\times 10^5 d_{kpc}^2$.  However, Figure \ref{bias2}, showing the maximum mass detected in each distance bin and the theoretical upper limit due to the limited field-of-view, demonstrates that the maximum mass allowed by the limited field-of-view of the GRS, $M_{max}^{FOV}$, is never reached. Thus, we use a constant upper bound, $M_{max}$, in the integration of Equation \ref{average_mass_eq1}. We use $M_{max}$ $=$ 10$^6$\Msun, the largest molecular cloud mass observed in the GRS, and also a typical high end on the mass of massive giant molecular clouds \citep{solomon87}.

\subsubsection{Biases observed in the GRS}
\begin{figure}
\centering
\includegraphics[width=\textwidth]{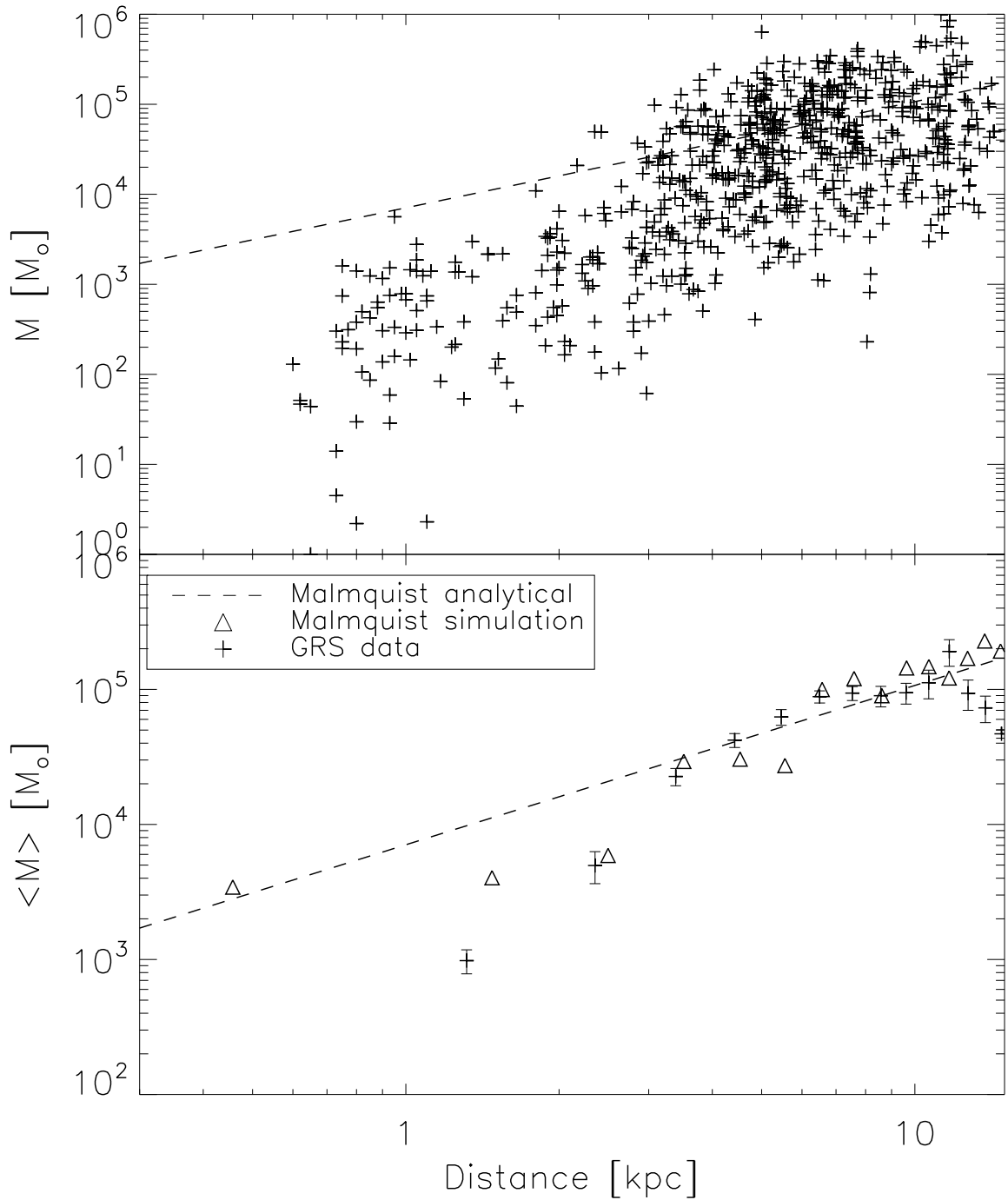}
\caption{Molecular clouds' individual masses and masses averaged within 1 kpc distance bins versus distance.  The dashed lines represent the expected trend from the analytical description of the Malmquist bias predicted for the GRS. The triangles are predictions from a numerical simulation of the Malmquist bias in the GRS.}
\label{bias1}
\end{figure}

\begin{figure}
\centering
\includegraphics[width=\textwidth]{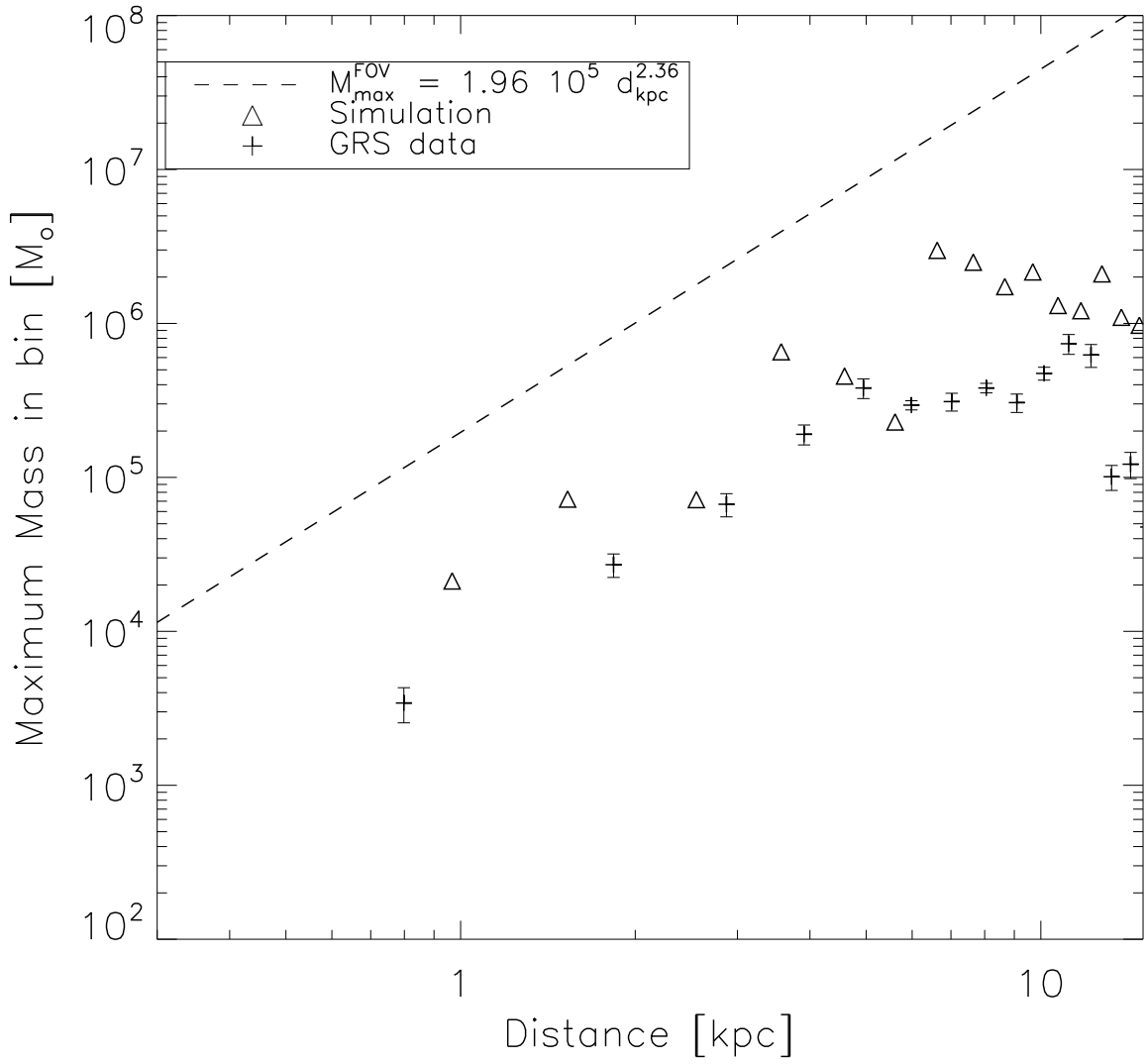}
\caption{Maximum molecular cloud mass detected in each 1 kpc distance bin as a function of distance. The dashed line represents the maximum mass allowed by the field-of-view of the GRS. }
\label{bias2}
\end{figure}

\indent Figure \ref{bias1} shows the molecular clouds' masses versus distance. The bottom panel shows the trend averaged over 1 kpc distance bins. The dashed line indicates the prediction from our simple analytical model of the Malmquist bias. Our analytical prediction for the Malmquist bias is not as steep as the observed trend, and does not reproduce the observed data well. This could be due to several effects. First, the probability of observing rare, massive clouds in the solar neighborhood is low due to the small volume probed at such a low distance, and due to the lack of massive Galactic structure in this area. As a result the high end of the mass spectrum is not well randomly sampled in our data. This tends to make the average molecular cloud mass at low distance lower than the prediction obtained from the Malmquist bias. Second, the surface density of  molecular gas, and thus the probability of observing massive molecular clouds, is not uniform in the Milky Way.\\
\indent To further investigate the cause of the observed distance/mass trend, we have performed a simple simulation. We have simulated a sample of molecular clouds with a mass spectrum $\phi(M) \propto M^{-1.64}$ that matches the Galactic surface mass density of molecular gas observed in Figure \ref{plot_rg_masses}. To that end, we have populated concentric rings of constant Galactocentric radius and of thickness 50 pc centered on the galactic center with molecular clouds, in order to match the observed Galactic surface density of molecular clouds. At each Galactocentric radius, the molecular cloud population in the ring is randomly sampled from a power-law mass spectrum of exponent $-$1.64, with masses ranging between 10 \Msu and 3$\times$10$^6$ \Msun. The azimuth of the clouds is randomly selected from a uniform distribution between 0 and 2$\pi$, and their distance and galactic longitude are computed based on their Galactocentric radius and azimuth. Molecular clouds that have masses greater than the completeness limit at their distance ($M_{min}$ $=$ 50 $d_{kpc}^2$) are included in the population of the ring, until the surface density of the ring reaches the observed surface density of molecular gas at this Galactocentric radius, given by Figure \ref{plot_rg_masses}. The simulation produced a sample of 2644 molecular clouds between  a Galactocentric radii 1.5 kpc and 10 kpc, 718 of which are located within the GRS longitude range. In this simulated sample of molecular clouds located inside the GRS coverage, we recorded the maximum and mean molecular cloud mass in each distance bin (we used 1 kpc distance bins), the same way we did for our GRS sample of molecular clouds. The results of this simulation is shown in Figure \ref{bias1} as triangles. \\
\indent Our simple simulation of the Malmquist bias seems to reproduce the observed trends within the error bars. We are therefore confident that the observed trends are indeed due to a Malmquist bias. Effects of the molecular cloud identification algorithm may play a secondary role in the relation between distance and molecular cloud masses at small distances. Indeed, CLUMPFIND tends to detect well resolved, nearby molecular clouds as distinct objects while the same molecular clouds at a farther distance would be blended together and assigned to a single cloud.

\subsubsection{Effects of biases on molecular clouds' physical properties}

\indent Biases mainly affect the significance of the radial profiles derived in Section \ref{rg_prop_section}. Because the masses of the molecular clouds on average increase with distance, the radial variations of the molecular gas content of the Milky Way seen in Figure \ref{plot_rg_masses}  would result from biases if the different radial bins were dominated by molecular clouds that were on average at different ranges of distance. The dash-dot lines in each panel of Figure \ref{plot_rg_masses} represent $<d>_{0.5kpc}$, the average of  $d$ in each 0.5 kpc galactocentric radius bin,  versus distance. Because the Malmquist bias causes the average molecular cloud mass to increase with distance, $<d>_{0.5kpc} (R_{\mathrm{gal}})$ should give us an indication of whether or not the radial profiles shown in Figure \ref{plot_rg_masses} are dominated by biases. There is no obvious correlation between $<d>_{0.5kpc}(R_{\mathrm{gal}})$ and $<M>(R_{\mathrm{ga}})$ or $<\Sigma_c(R_{gal})$, indicating that the variations of $<M>$ and $\Sigma_c$ with $R_{gal}$ are genuine.

\subsection{Effects of \CO abundance variations}\label{abundance}
\indent The derivation of the GRS molecular clouds' masses is based on the assumption that the abundance of \CO relative to H$_2$ is uniform within a molecular cloud. Owing to photo-dissociation and fractionation, the abundance ratio $n(^{13}\mathrm{CO})/n(\mathrm{H}_2)$ may however decrease significantly between the shielded dense interiors of molecular clouds and their diffuse, UV-exposed envelopes \citep{liszt07, glover10}. Molecular hydrogen (H$_2$) being self-shielded due to numerous optically thick absorption lines in the UV, it can resist the photo-dissociating ISRF at lower extinctions (A$_V$ of a few tenth) than its CO counterpart, which can only exist for A$_V$ $>$ 1-3 \citep{glover10, wolfire10}.  A significant amount of molecular gas is therefore likely hidden in CO dark envelopes around molecular clouds. As a result, our method probably underestimates the masses of the GRS molecular clouds. \citet{wolfire10} predicts that 30\% of a molecular cloud's mass is in the form of CO-dark molecular envelope, while \citet{goldsmith08} predicts 50\%. This is nonetheless the order of magnitude of the molecular mass not traced by CO in the Milky Way. \\
\indent In addition, the CO molecule tends to leave the gas phase and freeze out onto dust grains (thus forming a mantle) in cold, dense molecular cores. This effect could also potentially render some of the molecular mass inside a molecular cloud invisible to CO observations. Indeed,  using other molecules such as N$_2$H$^+$ and the dust continuum as a tracer for the densest molecular phase, \citet{bacmann02} have measured a CO under-abundance of 5-15 in molecular cores of density 10$^5$-10$^6$ cm$^{-3}$. The under-abundance depends on the density \citep{bacmann02}, such that CO manteling on dust grain does not occur for densities lower than 10$^4$ cm$^{-3}$. The typical size of such cores is 0.05-0.2 pc \citep{bacmann00, rathborne10}, which at a distance of 5 kpc respresents an angular size of 15''.  Thus, a core's projected area represents 1\% of the GRS beam area. Therefore, we do not expect CO manteling on dust grain to have a significant effect on the estimation of a molecular cloud's mass compared to photo-dissociation effects. 

\subsection{Effects of sub-thermal excitation} \label{subthermal_section}
\indent In order to derive \CO optical depths and molecular cloud masses, the CO excitation temperature was assumed to be identical for \CO and $^{12}$CO. Since CO is usually thermalized within molecular clouds due to its low dipole moment, this is a reasonable assumption. This assumption might however break down in the more diffuse envelopes of molecular clouds, where the optically thick \COT can remain thermalized due to radiative trapping, while the optically thin \CO is sub-thermally excited. In this case, the \CO excitation temperature would be lower than the \COT excitation temperature.  Nonetheless, emission from subthermally excited \CO in diffuse molecular cloud envelopes is probably under the GRS detection threshold \citep{heyer09}. 

\subsection{Uncertainties on the mass estimation}\label{errors_section}
\indent In addition to the effects of sub-thermal excitation and abundance variations, noise intrinsic to the data, and errors on  kinematic distances affect the accuracy of a molecular cloud's mass estimate. The estimation of the masses of the molecular clouds is based on the knowledge, at each voxel, of the excitation temperature (derived from the \COT brightness temperature), of the optical depth (derived from the excitation temperature and the \CO brightness temperature), and the distance of each molecular cloud. The \CO and \COT brightness temperatures are affected by gaussian noise, with a standard deviation of $\sigma_{T_{mb}} = 0.26$ K for both the UMSB ($^{12}$CO) and the GRS ($^{13}$CO) surveys \citep{GRS}. The accuracy of kinematic distances was discussed in \citet{RD2009}. The error on the distance stems from the error in the velocity of a molecular cloud with respect to its LSR velocity, the difference being caused by cloud-to-cloud dispersion and non-circular motions associated with spiral arms. The cloud-to-cloud dispersion amounts to 3-5 \kms \citep{C85} and local velocity perturbations associated with spiral arms are of order 15 \kms \citep{C85}. \\
\indent The error on the molecular clouds' masses due to each of those factors was estimated using Monte-Carlo simulations.  For every molecular cloud in the GRS, we computed the velocity dispersion  and peak brightness temperature of the \CO line toward each GRS pixel. The \CO line toward each pixel of the molecular cloud was then modeled by a gaussian line with the same peak brightness temperature and velocity dispersion. Gaussian noise of standard deviation $\sigma_{T_{mb}} = 0.26$ K was added to the model molecular cloud, and its \CO optical depth estimated assuming an excitation temperature of 6.3 K (the average value in Figure \ref{histos_excpar}).  We reproduced the contribution of the distance uncertainty on the mass error using the following method. First, a random gaussian velocity error $e_v$ of dispersion $\sigma_{e_v}$ = 5 \kms was added to the LSR velocity of the molecular cloud, $V_0$, to obtain a flawed estimate of the molecular cloud LSR velocity $V_1$ = $V_0$ + $e_v$. The kinematic distance $d_1$ corresponding to $V_1$ was computed using the \citet{C85} rotation curve and the Galactic longitude of the molecular cloud, conserving the near/far kinematic distance assignment of the molecular clouds from \citet{RD2009}. The distance $d_1$ was finally used, along with the \CO optical depth of the model molecular cloud, to compute its mass. This process was repeated 20 times for the standard deviation of the mass of the model molecular cloud to converge. The error on the mass estimation of the molecular cloud was then computed as  $\sigma_M$ $=$ $\sqrt{<(M_i-M_{\mathrm{cloud}})^2>_{i=0,20} }$, where $M_{\mathrm{cloud}}$ is the mass of the GRS molecular cloud, and $M_i$ is the mass of the i$^{th}$ realization of the model molecular cloud. Error bars on the mass estimate are specified in Table \ref{table1} and in the online complete version of the table.

\section{Conclusion}\label{conclusion}
\indent We have derived the physical properties (radius, mass, excitation temperature, optical depth, virial parameter, density, and surface density) of 580 molecular clouds identified in the GRS and covered by the UMSB \COT survey.  We have derived the histograms of these properties, and found a power-law decrease of exponent $-$1.64$\pm$0.25 for the mass spectrum of molecular clouds, consistent with previous results from the literature. The median virial parameter of nearly 0.5 suggests that molecular clouds are gravitationally bound entities. The range of values for the physical properties is entended to constrain numerial models of molecular cloud formation and evolution. \\
\indent We have found a tight power-law correlation of exponent 2.36$\pm$0.04 between the radii and masses of molecular clouds. Based on this correlation, we have deduced that the fractal dimension of the ISM must be of order 2.36 in the molecular phase, and in the range of spatial scales covered by molecular clouds. The correlation between molecular cloud radii and masses also allowed us to derive masses for an additional 170 GRS molecular clouds not covered by the UMSB \COT survey, for which excitation temperatures and optical depths could not be derived. Based on the 750 molecular clouds, we have examined the Galactic surface mass density of molecular clouds in the Galactic region covered by the GRS. The azimuthally averaged Galactic surface mass density of molecular gas is enhanced by a factor of 5 between galactocentric radii of 4 and 5 kpc. This supports previous observations that most of the molecular content of the Milky Way is contained in a ring located 4-5 kpc away from the Galactic center \citep{clemens88}.  In addition, the two-dimensional map of the Galactic surface mass density of molecular gas is enhanced along the positions of the Scutum-Crux and Sagittarius arms (also detected with other tracers such as star counts, \hiis regions, magnetic fields, electron density, etc ...). We have found no enhancement of molecular gas along the Perseus arm, which could be due to uncertainties in the kinematic distance, near/far blending, or the lack of completeness at this distance ($>$ 10 kpc). Nonetheless, the molecular gas enhancements observed at the assumed positions of the Scutum and Sagittarius arms may suggest that molecular clouds form in spiral arms, and are disrupted in the inter-arm space. \\
\indent Last, we have found that the CO excitation temperature of molecular clouds decreases with galactocentric radius, suggesting a decrease in the star formation rate away from the Galactic center. The excitation temperature of molecular clouds is also marginally enhanced at a galactocentric radius of 6 kpc, which, in the GRS longitude range, corresponds to the inferred position of the Sagittarius arm. This marginal increase may be related to star formation activity in the Sagittarius arm. \\

\acknowledgments{
This work was supported by NSF grant AST-0507657. The molecular line data used in this paper is from the Boston University (BU)-FCRAO GRS, a joint project of Boston University and the Five College Radio Astronomy observatory funded by the National Science Fundation under grants AST 98-00334, AST 00-98562, AST 01-00793, AST 02-28993, and AST 05-07657. }

\bibliographystyle{apj}
\bibliography{biblio1}

\begin{thebibliography}{79}
\expandafter\ifx\csname natexlab\endcsname\relax\def\natexlab#1{#1}\fi

\bibitem[{{Aharonian} {et~al.}(2008){Aharonian}, {Akhperjanian}, {Bazer-Bachi},
  {Behera}, {Beilicke}, {Benbow}, {Berge}, {Bernl{\"o}hr}, {Boisson}, {Bolz},
  {Borrel}, {Braun}, {Brion}, {Brown}, {B{\"u}hler}, {Bulik}, {B{\"u}sching},
  {Boutelier}, {Carrigan}, {Chadwick}, {Chounet}, {Clapson}, {Coignet},
  {Cornils}, {Costamante}, {Degrange}, {Dickinson}, {Djannati-Ata{\"i}},
  {Domainko}, {O'C.~Drury}, {Dubus}, {Dyks}, {Egberts}, {Emmanoulopoulos},
  {Espigat}, {Farnier}, {Feinstein}, {Fiasson}, {F{\"o}rster}, {Fontaine},
  {Fukui}, {Funk}, {Funk}, {F{\"u}{\ss}ling}, {Gallant}, {Giebels},
  {Glicenstein}, {Gl{\"u}ck}, {Goret}, {Hadjichristidis}, {Hauser}, {Hauser},
  {Heinzelmann}, {Henri}, {Hermann}, {Hinton}, {Hoffmann}, {Hofmann},
  {Holleran}, {Hoppe}, {Horns}, {Jacholkowska}, {de Jager}, {Kendziorra},
  {Kerschhaggl}, {Kh{\'e}lifi}, {Komin}, {Kosack}, {Lamanna}, {Latham}, {Le
  Gallou}, {Lemi{\`e}re}, {Lemoine-Goumard}, {Lenain}, {Lohse}, {Martin},
  {Martineau-Huynh}, {Marcowith}, {Masterson}, {Maurin}, {McComb}, {Moderski},
  {Moriguchi}, {Moulin}, {de Naurois}, {Nedbal}, {Nolan}, {Olive}, {Orford},
  {Osborne}, {Ostrowski}, {Panter}, {Pedaletti}, {Pelletier}, {Petrucci},
  {Pita}, {P{\"u}hlhofer}, {Punch}, {Ranchon}, {Raubenheimer}, {Raue},
  {Rayner}, {Reimer}, {Renaud}, {Ripken}, {Rob}, {Rolland}, {Rosier-Lees},
  {Rowell}, {Rudak}, {Ruppel}, {Sahakian}, {Santangelo}, {Saug{\'e}},
  {Schlenker}, {Schlickeiser}, {Schr{\"o}der}, {Schwanke}, {Schwarzburg},
  {Schwemmer}, {Shalchi}, {Sol}, {Spangler}, {Stawarz}, {Steenkamp},
  {Stegmann}, {Superina}, {Takeuchi}, {Tam}, {Tavernet}, {Terrier}, {van
  Eldik}, {Vasileiadis}, {Venter}, {Vialle}, {Vincent}, {Vivier}, {V{\"o}lk},
  {Volpe}, {Wagner}, \& {Ward}}]{aharonian08}
{Aharonian}, F., {et~al.} 2008, \aap, 481, 401

\bibitem[{{Allen}(1973)}]{allen73}
{Allen}, C.~W. 1973, {Astrophysical quantities}, ed. {Allen, C.~W.}

\bibitem[{{Anderson} {et~al.}(2009){Anderson}, {Bania}, {Jackson}, {Clemens},
  {Heyer}, {Simon}, {Shah}, \& {Rathborne}}]{anderson09}
{Anderson}, L.~D., {Bania}, T.~M., {Jackson}, J.~M., {Clemens}, D.~P., {Heyer},
  M., {Simon}, R., {Shah}, R.~Y., \& {Rathborne}, J.~M. 2009, \apjs, 181, 255

\bibitem[{{Bacmann} {et~al.}(2000){Bacmann}, {Andr{\'e}}, {Puget}, {Abergel},
  {Bontemps}, \& {Ward-Thompson}}]{bacmann00}
{Bacmann}, A., {Andr{\'e}}, P., {Puget}, J., {Abergel}, A., {Bontemps}, S., \&
  {Ward-Thompson}, D. 2000, \aap, 361, 555

\bibitem[{{Bacmann} {et~al.}(2002){Bacmann}, {Lefloch}, {Ceccarelli},
  {Castets}, {Steinacker}, \& {Loinard}}]{bacmann02}
{Bacmann}, A., {Lefloch}, B., {Ceccarelli}, C., {Castets}, A., {Steinacker},
  J., \& {Loinard}, L. 2002, \aap, 389, L6

\bibitem[{{Bakes} \& {Tielens}(1994)}]{bakes94}
{Bakes}, E.~L.~O., \& {Tielens}, A.~G.~G.~M. 1994, \apj, 427, 822

\bibitem[{{Bazell} \& {Desert}(1988)}]{bazell88}
{Bazell}, D., \& {Desert}, F.~X. 1988, \apj, 333, 353

\bibitem[{{Beech}(1987)}]{beech87}
{Beech}, M. 1987, \apss, 133, 193

\bibitem[{{Benjamin}(2009)}]{benjamin09}
{Benjamin}, R.~A. 2009, in IAU Symposium, Vol. 254, IAU Symposium, ed.
  {J.~Andersen, J.~Bland-Hawthorn, \& B.~Nordstr{\"o}m}, 319--322

\bibitem[{{Benjamin} {et~al.}(2005){Benjamin}, {Churchwell}, {Babler},
  {Indebetouw}, {Meade}, {Whitney}, {Watson}, {Wolfire}, {Wolff}, {Ignace},
  {Bania}, {Bracker}, {Clemens}, {Chomiuk}, {Cohen}, {Dickey}, {Jackson},
  {Kobulnicky}, {Mercer}, {Mathis}, {Stolovy}, \& {Uzpen}}]{benjamin05}
{Benjamin}, R.~A., {et~al.} 2005, \apjl, 630, L149

\bibitem[{{Bernard} \& {et al.}(2008)}]{bernard08}
{Bernard}, J., \& {et al.} 2008, \aj, 136, 919

\bibitem[{{Black} \& {Dalgarno}(1977)}]{black77}
{Black}, J.~H., \& {Dalgarno}, A. 1977, \apjs, 34, 405

\bibitem[{{Blake} {et~al.}(1987){Blake}, {Sutton}, {Masson}, \&
  {Phillips}}]{blake87}
{Blake}, G.~A., {Sutton}, E.~C., {Masson}, C.~R., \& {Phillips}, T.~G. 1987,
  \apj, 315, 621

\bibitem[{{Bloemen} {et~al.}(1986){Bloemen}, {Strong}, {Mayer-Hasselwander},
  {Blitz}, {Cohen}, {Dame}, {Grabelsky}, {Thaddeus}, {Hermsen}, \&
  {Lebrun}}]{bloemen86}
{Bloemen}, J.~B.~G.~M., {et~al.} 1986, \aap, 154, 25

\bibitem[{{Boulanger} {et~al.}(1996){Boulanger}, {Abergel}, {Bernard},
  {Burton}, {Desert}, {Hartmann}, {Lagache}, \& {Puget}}]{boulanger96}
{Boulanger}, F., {Abergel}, A., {Bernard}, J., {Burton}, W.~B., {Desert}, F.,
  {Hartmann}, D., {Lagache}, G., \& {Puget}, J. 1996, \aap, 312, 256

\bibitem[{{Brunt} {et~al.}(2010){Brunt}, {Federrath}, \& {Price}}]{brunt10}
{Brunt}, C.~M., {Federrath}, C., \& {Price}, D.~J. 2010, \mnras, 403, 1507

\bibitem[{{Burton} {et~al.}(1990){Burton}, {Hollenbach}, \&
  {Tielens}}]{burton90}
{Burton}, M.~G., {Hollenbach}, D.~J., \& {Tielens}, A.~G.~G.~M. 1990, \apj,
  365, 620

\bibitem[{{Churchwell} {et~al.}(2009){Churchwell}, {Babler}, {Meade},
  {Whitney}, {Benjamin}, {Indebetouw}, {Cyganowski}, {Robitaille}, {Povich},
  {Watson}, \& {Bracker}}]{churchwell09}
{Churchwell}, E., {et~al.} 2009, \pasp, 121, 213

\bibitem[{{Clemens}(1985)}]{C85}
{Clemens}, D.~P. 1985, \apj, 295, 422

\bibitem[{{Clemens} {et~al.}(1988){Clemens}, {Sanders}, \&
  {Scoville}}]{clemens88}
{Clemens}, D.~P., {Sanders}, D.~B., \& {Scoville}, N.~Z. 1988, \apj, 327, 139

\bibitem[{{Clemens} {et~al.}(1986){Clemens}, {Sanders}, {Scoville}, \&
  {Solomon}}]{clemens86}
{Clemens}, D.~P., {Sanders}, D.~B., {Scoville}, N.~Z., \& {Solomon}, P.~M.
  1986, \apjs, 60, 297

\bibitem[{{Dame} {et~al.}(1986){Dame}, {Elmegreen}, {Cohen}, \&
  {Thaddeus}}]{dame86}
{Dame}, T.~M., {Elmegreen}, B.~G., {Cohen}, R.~S., \& {Thaddeus}, P. 1986,
  \apj, 305, 892

\bibitem[{{Dobbs} {et~al.}(2006){Dobbs}, {Bonnell}, \& {Pringle}}]{dobbs06}
{Dobbs}, C.~L., {Bonnell}, I.~A., \& {Pringle}, J.~E. 2006, \mnras, 371, 1663

\bibitem[{{Elmegreen} \& {Falgarone}(1996)}]{elmegreen96}
{Elmegreen}, B.~G., \& {Falgarone}, E. 1996, \apj, 471, 816

\bibitem[{{Falgarone}(1989)}]{falgarone89}
{Falgarone}, E. 1989, in Lecture Notes in Physics, Berlin Springer Verlag, Vol.
  350, IAU Colloq. 120: Structure and Dynamics of the Interstellar Medium, ed.
  {G.~Tenorio-Tagle, M.~Moles, \& J.~Melnick}, 68--+

\bibitem[{{Falgarone} \& {Phillips}(1991)}]{falgarone91}
{Falgarone}, E., \& {Phillips}, T.~G. 1991, in IAU Symposium, Vol. 147,
  Fragmentation of Molecular Clouds and Star Formation, ed. {E.~Falgarone,
  F.~Boulanger, \& G.~Duvert}, 119--+

\bibitem[{{Federrath} {et~al.}(2009){Federrath}, {Klessen}, \&
  {Schmidt}}]{fed09a}
{Federrath}, C., {Klessen}, R.~S., \& {Schmidt}, W. 2009, \apj, 692, 364

\bibitem[{{Field} \& {Saslaw}(1965)}]{field65}
{Field}, G.~B., \& {Saslaw}, W.~C. 1965, \apj, 142, 568

\bibitem[{{Gabici} {et~al.}(2009){Gabici}, {Aharonian}, \&
  {Casanova}}]{gabici09}
{Gabici}, S., {Aharonian}, F.~A., \& {Casanova}, S. 2009, \mnras, 396, 1629

\bibitem[{{Glover} \& {Mac Low}(2010)}]{glover10}
{Glover}, S.~C.~O., \& {Mac Low}, M. 2010, \mnras, submitted, arXiv:1003.1340

\bibitem[{{Goldsmith} {et~al.}(2008){Goldsmith}, {Heyer}, {Narayanan}, {Snell},
  {Li}, \& {Brunt}}]{goldsmith08}
{Goldsmith}, P.~F., {Heyer}, M., {Narayanan}, G., {Snell}, R., {Li}, D., \&
  {Brunt}, C. 2008, \apj, 680, 428

\bibitem[{{Goldsmith} \& {Langer}(1978)}]{goldsmith78}
{Goldsmith}, P.~F., \& {Langer}, W.~D. 1978, \apj, 222, 881

\bibitem[{{Gordon} {et~al.}(2003){Gordon}, {Clayton}, {Misselt}, {Landolt}, \&
  {Wolff}}]{gordon03}
{Gordon}, K.~D., {Clayton}, G.~C., {Misselt}, K.~A., {Landolt}, A.~U., \&
  {Wolff}, M.~J. 2003, \apj, 594, 279

\bibitem[{{Gordon} {et~al.}(2010){Gordon}, {Galliano}, {Hony}, {Bernard},
  {Bolatto}, {Bot}, {Engelbracht}, {Hughes}, {Israel}, {Kemper}, {Kim}, {Li},
  {Madden}, {Matsuura}, {Meixner}, {Misselt}, {Okumura}, {Panuzzo}, {Rubio},
  {Reach}, {Roman-Duval}, {Sauvage}, {Skibba}, \& {Tielens}}]{gordon10}
{Gordon}, K.~D., {et~al.} 2010, \aap, 518, L89

\bibitem[{{Heitsch} {et~al.}(2001){Heitsch}, {Mac Low}, \&
  {Klessen}}]{heitsch01}
{Heitsch}, F., {Mac Low}, M., \& {Klessen}, R.~S. 2001, \apj, 547, 280

\bibitem[{{Heyer} {et~al.}(2009){Heyer}, {Krawczyk}, {Duval}, \&
  {Jackson}}]{heyer09}
{Heyer}, M., {Krawczyk}, C., {Duval}, J., \& {Jackson}, J.~M. 2009, \apj, 699,
  1092

\bibitem[{{Heyer} {et~al.}(2001){Heyer}, {Carpenter}, \& {Snell}}]{heyer01}
{Heyer}, M.~H., {Carpenter}, J.~M., \& {Snell}, R.~L. 2001, \apj, 551, 852

\bibitem[{{Heyer} \& {Terebey}(1998)}]{heyer98}
{Heyer}, M.~H., \& {Terebey}, S. 1998, \apj, 502, 265

\bibitem[{{Hollenbach} \& {McKee}(1989)}]{hollenbach89}
{Hollenbach}, D., \& {McKee}, C.~F. 1989, \apj, 342, 306

\bibitem[{{Jackson} {et~al.}(2006){Jackson}, {Rathborne}, {Shah}, {Simon},
  {Bania}, {Clemens}, {Chambers}, {Johnson}, {Dormody}, {Lavoie}, \&
  {Heyer}}]{GRS}
{Jackson}, J.~M., {et~al.} 2006, \apjs, 163, 145

\bibitem[{{Kerr} \& {Lynden-Bell}(1986)}]{kerr86}
{Kerr}, F.~J., \& {Lynden-Bell}, D. 1986, \mnras, 221, 1023

\bibitem[{{Klessen} {et~al.}(2000){Klessen}, {Heitsch}, \& {Mac
  Low}}]{klessen00}
{Klessen}, R.~S., {Heitsch}, F., \& {Mac Low}, M. 2000, \apj, 535, 887

\bibitem[{{Langer} \& {Penzias}(1990)}]{langer90}
{Langer}, W.~D., \& {Penzias}, A.~A. 1990, \apj, 357, 477

\bibitem[{{Larson}(1981)}]{larson81}
{Larson}, R.~B. 1981, \mnras, 194, 809

\bibitem[{{Liszt}(2006)}]{liszt06}
{Liszt}, H.~S. 2006, \aap, 447, 533

\bibitem[{{Liszt}(2007)}]{liszt07}
---. 2007, \aap, 476, 291

\bibitem[{{Mandelbrot} \& {Whitrow}(1983)}]{mandelbrot83}
{Mandelbrot}, B.~B., \& {Whitrow}, G.~J. 1983, Journal of the British
  Astronomical Association, 93, 238

\bibitem[{{McKee}(1989)}]{mckee89}
{McKee}, C.~F. 1989, \apj, 345, 782

\bibitem[{{Meneveau} \& {Sreenivasan}(1990)}]{meneveau89}
{Meneveau}, C., \& {Sreenivasan}, K.~R. 1990, \pra, 41, 2246

\bibitem[{{Oort}(1954)}]{oort54}
{Oort}, J.~H. 1954, \bain, 12, 177

\bibitem[{{Quireza} {et~al.}(2006){Quireza}, {Rood}, {Bania}, {Balser}, \&
  {Maciel}}]{quireza06}
{Quireza}, C., {Rood}, R.~T., {Bania}, T.~M., {Balser}, D.~S., \& {Maciel},
  W.~J. 2006, \apj, 653, 1226

\bibitem[{{Rathborne} {et~al.}(2010){Rathborne}, {Jackson}, {Chambers},
  {Stojimirovic}, {Simon}, {Shipman}, \& {Frieswijk}}]{rathborne10}
{Rathborne}, J.~M., {Jackson}, J.~M., {Chambers}, E.~T., {Stojimirovic}, I.,
  {Simon}, R., {Shipman}, R., \& {Frieswijk}, W. 2010, \apj, 715, 310

\bibitem[{{Rathborne} {et~al.}(2009){Rathborne}, {Johnson}, {Jackson}, {Shah},
  \& {Simon}}]{rathborne08}
{Rathborne}, J.~M., {Johnson}, A.~M., {Jackson}, J.~M., {Shah}, R.~Y., \&
  {Simon}, R. 2009, \apjs, 182, 131

\bibitem[{{Reid} {et~al.}(2009){Reid}, {Menten}, {Zheng}, {Brunthaler},
  {Moscadelli}, {Xu}, {Zhang}, {Sato}, {Honma}, {Hirota}, {Hachisuka}, {Choi},
  {Moellenbrock}, \& {Bartkiewicz}}]{reid09}
{Reid}, M.~J., {et~al.} 2009, \apj, 700, 137

\bibitem[{{Rohlfs} \& {Wilson}(2004)}]{rohlfs03}
{Rohlfs}, K., \& {Wilson}, T.~L. 2004, {Tools of radio astronomy}, ed. {Rohlfs,
  K.~\& Wilson, T.~L.}

\bibitem[{{Roman-Duval} {et~al.}(in preparation){Roman-Duval}, {Federrath},
  {Heyer}, {Jackson}, {Klessen}, \& {Mac Low}}]{rd10}
{Roman-Duval}, J., {Federrath}, C., {Heyer}, M., {Jackson}, J., {Klessen}, R.,
  \& {Mac Low}, M.-M. in preparation

\bibitem[{{Roman-Duval} {et~al.}(2009){Roman-Duval}, {Jackson}, {Heyer},
  {Johnson}, {Rathborne}, {Shah}, \& {Simon}}]{RD2009}
{Roman-Duval}, J., {Jackson}, J.~M., {Heyer}, M., {Johnson}, A., {Rathborne},
  J., {Shah}, R., \& {Simon}, R. 2009, \apj, 699, 1153

\bibitem[{{Sanders} {et~al.}(1986){Sanders}, {Clemens}, {Scoville}, \&
  {Solomon}}]{sanders86}
{Sanders}, D.~B., {Clemens}, D.~P., {Scoville}, N.~Z., \& {Solomon}, P.~M.
  1986, \apjs, 60, 1

\bibitem[{{Sanders} {et~al.}(1985){Sanders}, {Scoville}, \&
  {Solomon}}]{sanders85}
{Sanders}, D.~B., {Scoville}, N.~Z., \& {Solomon}, P.~M. 1985, \apj, 289, 373

\bibitem[{{Scalo}(1990)}]{scalo90}
{Scalo}, J. 1990, in Astrophysics and Space Science Library, Vol. 162, Physical
  Processes in Fragmentation and Star Formation, ed. {R.~Capuzzo-Dolcetta,
  C.~Chiosi, \& A.~di Fazio}, 151--176

\bibitem[{{Scalo}(1985)}]{scalo85}
{Scalo}, J.~M. 1985, in Protostars and Planets II, ed. {D.~C.~Black \&
  M.~S.~Matthews}, 201--296

\bibitem[{{Scalo}(1988)}]{scalo88}
{Scalo}, J.~M. 1988, in Lecture Notes in Physics, Berlin Springer Verlag, Vol.
  315, Molecular Clouds, Milky-Way and External Galaxies, ed. {R.~L.~Dickman,
  R.~L.~Snell, \& J.~S.~Young}, 201

\bibitem[{{Simon} {et~al.}(2001){Simon}, {Jackson}, {Clemens}, {Bania}, \&
  {Heyer}}]{simon01}
{Simon}, R., {Jackson}, J.~M., {Clemens}, D.~P., {Bania}, T.~M., \& {Heyer},
  M.~H. 2001, \apj, 551, 747

\bibitem[{{Sodroski} {et~al.}(1997){Sodroski}, {Odegard}, {Arendt}, {Dwek},
  {Weiland}, {Hauser}, \& {Kelsall}}]{sodroski97}
{Sodroski}, T.~J., {Odegard}, N., {Arendt}, R.~G., {Dwek}, E., {Weiland},
  J.~L., {Hauser}, M.~G., \& {Kelsall}, T. 1997, \apj, 480, 173

\bibitem[{{Solomon} {et~al.}(1987){Solomon}, {Rivolo}, {Barrett}, \&
  {Yahil}}]{solomon87}
{Solomon}, P.~M., {Rivolo}, A.~R., {Barrett}, J., \& {Yahil}, A. 1987, \apj,
  319, 730

\bibitem[{{Sreenivasan}(1991)}]{sreenivasan91}
{Sreenivasan}, K.~R. 1991, Annual Review of Fluid Mechanics, 23, 539

\bibitem[{{Sreenivasan} \& {Meneveau}(1986)}]{sreenivasan86}
{Sreenivasan}, K.~R., \& {Meneveau}, C. 1986, Journal of Fluid Mechanics, 173,
  357

\bibitem[{{Strong} {et~al.}(1988){Strong}, {Bloemen}, {Dame}, {Grenier},
  {Hermsen}, {Lebrun}, {Nyman}, {Pollock}, \& {Thaddeus}}]{strong88}
{Strong}, A.~W., {et~al.} 1988, \aap, 207, 1

\bibitem[{{Tasker} \& {Tan}(2009)}]{tasker09}
{Tasker}, E.~J., \& {Tan}, J.~C. 2009, \apj, 700, 358

\bibitem[{{Urquhart} {et~al.}(2010){Urquhart}, {Moore}, {Hoare}, {Lumsden},
  {Oudmaijer}, {Rathborne}, {Mottram}, {Davies}, \& {Stead}}]{urquhart10}
{Urquhart}, J., {et~al.} 2010

\bibitem[{{Vallee}(1995)}]{vallee95}
{Vallee}, J.~P. 1995, \apj, 454, 119

\bibitem[{{Wakker}(1990)}]{wakker90}
{Wakker}, B.~P. 1990, PhD thesis, PhD thesis, Univ.~Groningen, (1990)

\bibitem[{{Whittet}(2003)}]{whittet03}
{Whittet}, D. 2003, Lunar and planetary information bulletin, no.~94, p.~11
  (2003), 94, 11

\bibitem[{{Williams} {et~al.}(2000){Williams}, {Blitz}, \&
  {McKee}}]{williams00}
{Williams}, J.~P., {Blitz}, L., \& {McKee}, C.~F. 2000, Protostars and Planets
  IV, 97

\bibitem[{{Williams} {et~al.}(1994){Williams}, {de Geus}, \&
  {Blitz}}]{williams94}
{Williams}, J.~P., {de Geus}, E.~J., \& {Blitz}, L. 1994, \apj, 428, 693

\bibitem[{{Williams} \& {McKee}(1997)}]{williams97}
{Williams}, J.~P., \& {McKee}, C.~F. 1997, \apj, 476, 166

\bibitem[{{Wolfire} {et~al.}(2010){Wolfire}, {Hollenbach}, \&
  {McKee}}]{wolfire10}
{Wolfire}, M.~G., {Hollenbach}, D., \& {McKee}, C.~F. 2010, \apj, 716, 1191

\bibitem[{{Wolfire} {et~al.}(2003){Wolfire}, {McKee}, {Hollenbach}, \&
  {Tielens}}]{wolfire03}
{Wolfire}, M.~G., {McKee}, C.~F., {Hollenbach}, D., \& {Tielens}, A.~G.~G.~M.
  2003, \apj, 587, 278

\bibitem[{{Zimmermann} \& {Stutzki}(1992)}]{zimmermann92}
{Zimmermann}, T., \& {Stutzki}, J. 1992, Physica A Statistical Mechanics and
  its Applications, 191, 79

\end{thebibliography}

\begin{deluxetable}{cccccccccccccc}
\centering
\tabletypesize{\tiny}
\tablecaption{Catalog of molecular clouds's masses and physical properties. \label{table1}}
\tablewidth{0pt}
\tablecolumns{14}
\tablehead{
\colhead{GRS molecular cloud} &
\colhead{l} &
\colhead{b} &
\colhead{$V_{LSR}$} &
\colhead{$\Delta V$}&
\colhead{$R$}&
\colhead{M} &
\colhead{$\delta$ M}&
\colhead{n(H$_2$)} &
\colhead{T$_{ex}$} &
\colhead{$\tau (^{13}CO)$} &
\colhead{$\Sigma_c$}&
\colhead{$\alpha_{vir}$} &
\colhead{}\\
\colhead{} &
\colhead{\tiny{(\degs)}} &
\colhead{\tiny{(\degs)}} &
\colhead{\tiny{(\kms)}}&
\colhead{\tiny{(\kms)}}&
\colhead{\tiny{(pc)}}&
\colhead{\tiny{(\Msun)}}&
\colhead{\tiny{(\Msun)}}&
\colhead{\tiny{(cm$^{-3}$)}}&
\colhead{\tiny{(K)}}&
\colhead{}&
\colhead{\tiny{(\Msu pc$^{-2}$)}}&
\colhead{}&
\colhead{}\\
 }
\startdata

GRSMC G053.59$+$00.04 & 53.59 &  0.04 &  23.74 &  1.86 &   17.5 & 2.35  $\times$ 10$^5$ &  0.566  $\times$ 10$^5$ &      161.0 &    8.28 &  1.78 &      244.4 &       0.04 & i\\
GRSMC G049.49$-$00.41 & 49.49 & -0.41 &  56.90 &  9.12 &    9.8 & 1.81  $\times$ 10$^5$ & 0.445  $\times$ 10$^5$ &      707.7 &    9.97 &  1.47 &      601.6 &       0.73 & i\\
GRSMC G018.89$-$00.51 & 18.89 & -0.51 &  65.82 &  2.80 &   12.4 & 1.41   $\times$ 10$^5$ & 0.457  $\times$ 10$^5$ &      271.9 &    9.72 &  0.89 &      292.4 &       0.11 & i\\
GRSMC G030.49$-$00.36 & 30.49 & -0.36 &  12.26 &  4.56 &    1.7 & 7.82  $\times$ 10$^5$ & 3.72  $\times$ 10$^2$ &      617.7 &    5.97 &  1.82 &       89.3 &       7.25 & i\\
GRSMC G035.14$-$00.76 & 35.14 & -0.76 &  35.22 &  4.89 &    5.4 & 4.95   $\times$ 10$^5$ & 1.21  $\times$ 10$^4$ &     1175.8 &    8.69 &  2.21 &      548.0 &       0.42 & i\\
GRSMC G034.24$+$00.14 & 34.24 &  0.14 &  57.75 &  5.66 &   12.3 & 1.44  $\times$ 10$^5$ &  0.406  $\times$ 10$^5$ &      281.0 &    6.48 &  1.83 &      300.6 &       0.45 & i\\
GRSMC G019.94$-$00.81 & 19.94 & -0.81 &  42.87 &  2.62 &    8.2 & 5.54   $\times$ 10$^5$ & 1.64  $\times$ 10$^4$ &      366.9 &    8.90 &  0.86 &      261.8 &       0.17 & i\\
GRSMC G038.94$-$00.46 & 38.94 & -0.46 &  41.59 &  2.78 &   22.8 & 4.88   $\times$ 10$^5$ & 1.03  $\times$ 10$^5$ &      150.2 &    8.27 &  1.77 &      297.8 &       0.06 & i\\
GRSMC G053.14$+$00.04 & 53.14 &  0.04 &  22.04 &  2.26 &    4.0 & 1.09  $\times$ 10$^5$ &  0.397  $\times$ 10$^4$ &      651.0 &    7.05 &  2.00 &      222.9 &       0.30 & i\\
GRSMC G022.44$+$00.34 & 22.44 &  0.34 &  84.52 &  2.62 &    7.5 & 5.22   $\times$ 10$^5$ & 1.20  $\times$ 10$^4$ &      448.2 &    7.99 &  1.29 &      293.3 &       0.16 & i\\
GRSMC G049.39$-$00.26 & 49.39 & -0.26 &  50.94 &  3.51 &   12.3 & 1.57  $\times$ 10$^5$ &  0.409  $\times$ 10$^5$ &      309.6 &    9.17 &  1.56 &      330.2 &       0.16 & i\\
GRSMC G019.39$-$00.01 & 19.39 & -0.01 &  26.72 &  3.63 &    6.3 & 4.93  $\times$ 10$^5$ &  1.85  $\times$ 10$^4$ &      732.0 &    7.11 &  2.05 &      399.1 &       0.27 & i\\
GRSMC G034.74$-$00.66 & 34.74 & -0.66 &  46.69 &  4.09 &    8.3 & 9.77  $\times$ 10$^5$ &  2.72  $\times$ 10$^4$ &      629.5 &    8.30 &  1.20 &      452.3 &       0.23 & i\\
GRSMC G023.04$-$00.41 & 23.04 & -0.41 &  74.32 &  4.14 &   15.8 & 2.23  $\times$ 10$^5$ &  0.679  $\times$ 10$^5$ &      208.9 &    9.16 &  0.92 &      285.6 &       0.20 & i\\
GRSMC G018.69$-$00.06 & 18.69 & -0.06 &  45.42 &  3.85 &   21.8 & 4.78   $\times$ 10$^5$ & 1.24  $\times$ 10$^5$ &      169.6 &    7.78 &  1.52 &      320.7 &       0.11 & i\\
GRSMC G023.24$-$00.36 & 23.24 & -0.36 &  77.30 &  2.65 &   15.5 & 1.95  $\times$ 10$^5$ &  0.683  $\times$ 10$^5$ &      190.9 &    8.64 &  1.01 &      257.3 &       0.09 & i\\
GRSMC G019.89$-$00.56 & 19.89 & -0.56 &  44.14 &  3.68 &    8.9 & 8.63   $\times$ 10$^5$ & 2.82  $\times$ 10$^4$ &      446.3 &    7.83 &  1.21 &      345.3 &       0.23 & i\\
GRSMC G022.04$+$00.19 & 22.04 &  0.19 &  50.94 &  5.53 &   10.8 & 8.48  $\times$ 10$^5$ &  1.93  $\times$ 10$^4$ &      248.3 &    7.17 &  1.15 &      232.3 &       0.63 & i\\
GRSMC G018.89$-$00.66 & 18.89 & -0.66 &  64.12 &  3.66 &   12.6 & 1.59  $\times$ 10$^5$ &  0.404  $\times$ 10$^5$ &      294.5 &    8.90 &  0.97 &      320.8 &       0.17 & i\\
GRSMC G023.34$-$00.21 & 23.34 & -0.21 &  81.12 &  4.09 &   13.1 & 1.63  $\times$ 10$^5$ &  0.509  $\times$ 10$^5$ &      267.4 &    8.87 &  0.97 &      303.6 &       0.22 & i\\
GRSMC G034.99$+$00.34 & 34.99 &  0.34 &  53.07 &  3.57 &   11.8 & 1.28  $\times$ 10$^5$ &  0.352  $\times$ 10$^5$ &      289.6 &    7.35 &  1.41 &      295.0 &       0.19 & i\\
GRSMC G029.64$-$00.61 & 29.64 & -0.61 &  75.60 &  3.79 &   13.1 & 9.58   $\times$ 10$^5$ & 3.01  $\times$ 10$^4$ &      156.9 &    6.61 &  1.37 &      178.2 &       0.32 & i\\
GRSMC G018.94$-$00.26 & 18.94 & -0.26 &  64.55 &  2.72 &   10.4 & 1.17  $\times$ 10$^5$ &  0.402  $\times$ 10$^5$ &      381.0 &    8.79 &  1.27 &      344.2 &       0.11 & i\\
GRSMC G024.94$-$00.16 & 24.94 & -0.16 &  47.12 &  4.40 &    9.5 & 9.21 $\times$ 10$^5$ &   2.14  $\times$ 10$^4$ &      398.5 &    8.57 &  1.00 &      327.5 &       0.32 & i\\
GRSMC G025.19$-$00.26 & 25.19 & -0.26 &  63.70 &  2.30 &    5.0 & 1.73  $\times$ 10$^5$ &  0.438  $\times$ 10$^4$ &      501.7 &    7.90 &  1.48 &      218.4 &       0.25 & i\\

\enddata
\label{mass_subtable}
\end{deluxetable}

\end{document}